\documentclass[a4paper, 11pt, oneside]{Thesis}  
\graphicspath{Figures/}  
\usepackage[square, numbers, comma, sort&compress]{natbib}  
\usepackage{verbatim}  
\usepackage{vector}  

\usepackage{algpseudocode}
\usepackage{multirow}
\usepackage{longtable}
\usepackage[linesnumbered,ruled,vlined]{algorithm2e}
\usepackage{mathtools}
\DeclarePairedDelimiter{\ceil}{\lceil}{\rceil}

\hypersetup{urlcolor=blue, colorlinks=true}  

\usepackage{fancyhdr}
\begin{document}
\frontmatter      

\UNIVERSITY{{THE UNIVERSITY OF MELBOURNE }}    
%
\department{{}}
\school{{School of Computing and Information Systems}}

%
\title  {FogBus2: A Lightweight and Distributed Container-based Framework for Integration of IoT-enabled Systems with Edge and Cloud Computing}
\authors  {\texorpdfstring
            {\href{qifand@student.unimelb.edu.au}{Qifan Deng }}
            {Qifan Deng} \\
            {\small Supervised under Prof. Rajkumar Buyya}
            }
\addresses  {\groupname\\\deptname\\\univname}  
\date       {\today}
\subject    {}

\maketitle

\setstretch{1.3}  

\fancyhead{}  
\rhead{\thepage}  
\lhead{}  

\pagestyle{fancy}  


\addtotoc{Abstract}  
\abstract{
\addtocontents{toc}{\vspace{1em}}  

Edge/Fog computing is a novel computing paradigm that provides resource-limited Internet of Things (IoT) devices with scalable computing and storage resources. Compared to cloud computing, edge/fog servers have fewer resources, but they can be accessed with higher bandwidth and less communication latency. Thus, integrating edge/fog and cloud infrastructures can support the execution of diverse latency-sensitive and computation-intensive IoT applications. Although some frameworks attempt to provide such integration, there are still several challenges to be addressed, such as dynamic scheduling of different IoT applications, scalability mechanisms, multi-platform support, and supporting different interaction models. To overcome these challenges, we propose a lightweight and distributed container-based framework, called FogBus2. It provides a mechanism for scheduling heterogeneous IoT applications and implements several scheduling policies. Also, it proposes an optimized genetic algorithm to obtain fast convergence to well-suited solutions. Besides, it offers a scalability mechanism to ensure efficient responsiveness when either the number of IoT devices increases or the resources become overburdened. Also, the dynamic resource discovery mechanism of FogBus2 assists new entities to quickly join the system. We have also developed two IoT applications, called Conway's Game of Life and Video Optical Character Recognition to demonstrate the effectiveness of FogBus2 for handling real-time and non-real-time IoT applications. Experimental results show FogBus2's scheduling policy improves the response time of IoT applications by 53\% compared to other policies. Also, the scalability mechanism can reduce up to 48\% of the queuing waiting time compared to frameworks that do not support scalability.
}
\clearpage  



\setstretch{1.3}  
\acknowledgements{
\addtocontents{toc}{\vspace{1em}}  

I would like firstly to thank my parents for their unconditional support. Days after I shared my consideration of continuing study in university, they sold their house without any hesitation, which costed their lifelong savings to cover my tuition and living costs during my Master's life. No word can express my gratitude for their giving and love.

Thanks to Professor Rajkumar Buyya, who guided me from the first month I arrived at university. I still remember the day I rashly ran to him, trying to get a chance to do my research under his supervision. Prof. Buyya kindly guided me to do well in the subject first and promised a meeting with me at the end of the semester. Prof. Buyya is highly experienced, and his professional and experienced guidance grow me from a fresh student to an expert in my research domain.

I also thank friends in the CLOUDS lab. They are willing to help and share their research and ideas, which I have learned a lot. Especial thanks to Mr. Mohammad Goudarzi. He is very experienced and gave me much advice in research which prevent me from going astray in the researching maze. 

Thanks to my considerate and pretty fiancée Siying. We have known each other for 4,165 days and have been together for 1,002 days. During these one thousand days, she supports me and helps me focus on the study and research with her understanding, patience, consideration, and love.

The environment on this planet of recent two years is unusual. But the worldwide collaboration and applied technologies are helping human beings get over the difficulties together, which convinces the importance of research, technology development, system development, empathy, and communication. I hope my future work can also contribute to people's efficiency and creativity, leaving my contribution a small footprint in human civilization's progress.

\vline

Qifan Deng \\
Minhang, Shanghai, China \\
Early morning, 4th June 2021 \\

}
\clearpage  

\pagestyle{fancy}  

\lhead{\emph{Contents}}  
\tableofcontents  

\lhead{\emph{List of Figures}}  
\listoffigures  

\lhead{\emph{List of Tables}}  
\listoftables  





\clearpage  
\lhead{\emph{\chaptername\ \thechapter\ --\ \leftmark}}  

\mainmatter	  
\pagestyle{fancy}  


\chapter{Introduction}
\label{chapt:introduction}

\begin{figure}[t]
    \centering 
    \includegraphics[width=\linewidth]{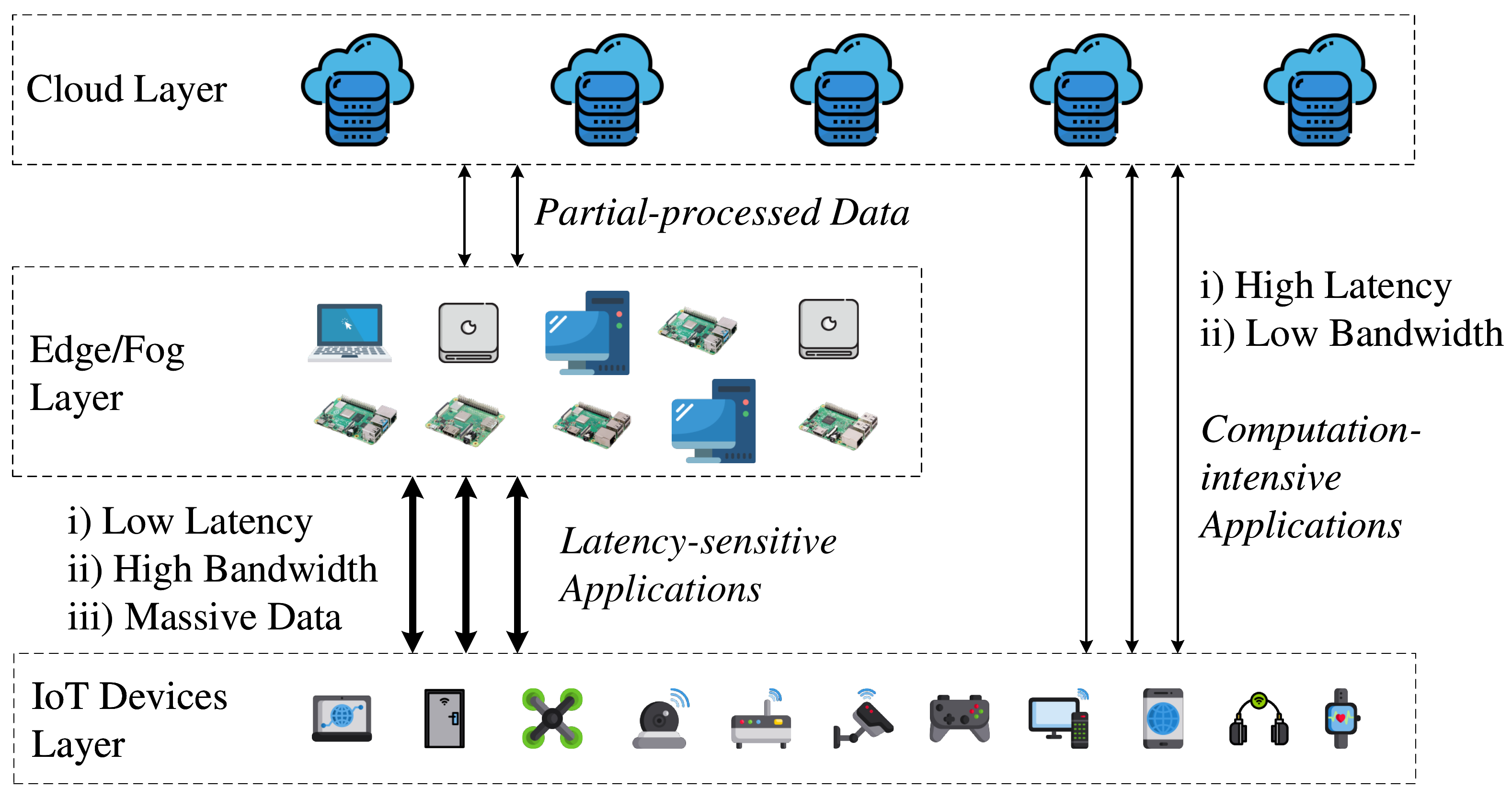}
    \caption{Environments and applications of IoT, Edg/fog and Cloud layers}
    \label{fig:iotAppAndEnv}
\end{figure}

Internet of Things (IoT) devices has become an inseparable part of our daily lives, where IoT applications provide diverse solutions for intelligent healthcare, transportation, and entertainment, to mention a few \cite{hu2017survey}. IoT applications often produce a massive amount of data for processing and storage. IoT devices connect to different networks with numerous sensors interacting with or sensing the internal and external environments on every second \cite{cook2019anomaly}. However, the computing and storage resources of IoT devices are limited. Therefore, IoT devices are usually integrated with resourceful surrogate resource providers to obtain better services for their users. Cloud computing, as a centralized computing paradigm, is one of the main enablers of IoT that offers unlimited computing and storage resources \cite{deng2020optimal,goudarzi2016mobile}. IoT devices can place whole or some parts of their applications to cloud servers for processing and storage. These applications benefit from cloud computing as the computing is delivered in the form of services provided by the cloud, which are stable and with high performance, \cite{5381983,furht2010handbook,mauch2013high}.

However, the networking between IoT devices and cloud clusters or data centers (which are usually aggregated) involves transportation of massive data over the Internet. This physical structure brings high latency when IoT applications use cloud computing as a part of the primary computing module in the design. The emergence of real-time IoT applications indicates that cloud computing cannot solely provide efficient services for latency-sensitive IoT applications due to its high access latency and low bandwidth \cite{xu2019computation,goudarzi2020application}. Moreover, although cloud services providers provide a `pay as you go' plan \cite{5192685,5329110,BELOGLAZOV2012755} that charges only on-demand computing, storage, and networking resources, it is expensive when the amount of data needed to be processed, transported and stored is enormous \cite{BAKER201796}.
To address these issues, edge/fog computing, which is a novel distributed computing paradigm, is proposed, providing distributed computing and storage resources in the proximity of IoT devices with higher access bandwidth and lower communication latency \cite{GOUDARZI2019102407}. Edge computing exists closely to IoT devices where data is generated. It computes, stores, and forms networks to serve IoT devices with low latency, high bandwidth, and distributed mobility \cite{8030322,7807196,7488250}. Edge servers process and analyzes data around where it is generated, the overload of transmission to a centralized data center is dramatically reduced. Combining with techniques like 5G networks, edge computing further enables faster data processing which boosts the evaluation of applications like autonomous electric transport vehicles, autonomous drones, healthcare devices, and autonomous robots in factories \cite{MISTRY2020106382,8519960}. 
The environment and applications of IoT, Edg/fog and Cloud layers are presented in Figure~\ref{fig:iotAppAndEnv}.

\section{Challenges and Motivation}
\label{sec:challengesAndMotivation}
Compared to cloud servers' resources, edge/fog servers have limited computing and storage resources, and hence they cannot efficiently execute computation-intensive tasks of IoT devices. To address this issue, edge/fog servers can collaboratively use their resources or cloud servers'. However, there is overload when simultaneously using distributed resources. Not only the algorithms and communication protocols need to be carefully designed, but networking bandwidth and latency also need to be considered. Thus, seamless integration of edge/fog and cloud infrastructures to support different IoT applications is an important research topic.
 Resources of distributed edge/fog servers and cloud servers are highly heterogeneous in terms of computing capabilities, processors' architectures, RAM capacity, and supported communication protocols \cite{merlino2019enabling}. Also, IoT applications are heterogeneous in terms of applications' granularity (i.e., task, service), dependency model of constituent parts of IoT applications (i.e., independent tasks, sequential dependency, and complex dependent tasks), and their quality of service requirements (such as computation-intensive or latency-sensitive applications).

According to these factors, there are several framework design challenges to be considered. First, frameworks working in the integrated platform should support platform-independent techniques to overcome communication and run-time obstacles. Second, due to the heterogeneity of resources and the requirements of IoT applications, distributed scheduling mechanisms are required to place/offload tasks/data of IoT applications on suitable servers for processing and storage. Third, fast application deployments and scalability support are required in this integrated environment to provide services for IoT devices in a timely manner. Fourth, to efficiently reuse the resources, the containerization concepts can be adopted for the software components of the framework and IoT applications.

\section{Research Problems}
\label{sec:researchProblems}
To address the challenges we discussed, the following questions are identified and investigated in this thesis. By answering these questions with solutions in the framework design, the challenges in the previous chapter should be completed.
\begin{itemize}
 \item \textbf{How to use edge/fog servers and cloud servers simultaneously in an efficient way?} When cloud servers are centralized, and with high performance, the access latency is always high. Thus, cloud servers cannot solely provide efficient services for latency-sensitive applications. Edge/fog servers are with lower performance but very high bandwidth and low latency to IoT devices complete cloud servers and offset the drawbacks of cloud servers. An efficient framework has to use both the resources of edge/fog servers and cloud servers collaboratively.
 \item \textbf{How to provide a scheduling mechanism for incoming requests from different types of IoT applications (latency-sensitive and computation-intensive)?} When the performance and resources diverse in the heterogeneous edge/fog servers and cloud servers, numerous applications may be latency-sensitive and or computation-intensive. A platform needs to intelligently decide how to arrange the executions for all the kind of applications. Typically, there is a time limitation for the scheduling algorithm to perform, which requires the algorithm to finish in a short time but finishes with a reasonable solution of the arrangement of the execution.
 \item \textbf{How to provide a scalable platform in these heterogeneous computing environments?} With the scheduling mechanism, a framework allows real-time scheduling for the heterogeneous application execution request. However, the demands need to be rapidly responded even there are a huge amount of requests coming to the system at the same time. To respond to the requests efficiently, scalability is required for frameworks that automatically scale, create, or allocate resources responding to the execution requests of IoT applications. This mechanism helps reduce user-side waiting time for resource placement, thus bring users a better experience.
 \item \textbf{How to support automatically discover resources and reuse for such platform?} The amount of IoT devices and edge/fog servers is countless and keep growing all the time. Under this background, a framework should have the ability to automatically identify resources when IoT devices and edge/fog servers join or leave the system unpredictably. Nevertheless, after discovering resources, an effective mechanism should be designed to operate the resources reasonably, including the reuse of the resources. Frameworks should recognize this problem during the design phase to better advantage all the heterogeneity IoT, edge/fog servers, and cloud servers.
\end{itemize}

\section{Thesis Contributions}
Although there are some frameworks to manage integrated resources in edge/fog computing \cite{tuli2019fogbus,an2019eif}, they barely consider platform-independent techniques, scheduling of heterogeneous IoT applications with complex dependent structures, scalability mechanisms of distributed resource managers, and containerization (see Chapter~\ref{chapt:relatedWork}).

To address these limitations and solve these research problems, we propose and develop a lightweight and distributed container-based framework called FogBus2, which is partially published in \cite{deng2021fogbus2}. Our framework supports (1) different inter and intra interaction models among edge/fog servers and cloud servers to support the requirements of different IoT application scenarios when using IP and ports addresses, (2) containerization of software components of the framework for fast deployments when supporting different programming languages by following the communication protocol design, (3) containerization of constituent parts of IoT applications as dependent tasks or independent tasks, (4) scheduling of multiple IoT applications and scalability mechanisms based on monitoring the whole system by integrating logs from every component (5) concurrent execution of different types of IoT applications including computation-intensive and latency-sensitive as well as complex dependent or simply related modules, and (6) efficient reuse of resources.

The main contributions of this paper are summarized as follows:
\begin{itemize}
\item A lightweight and distributed container-based framework, called FogBus2, is proposed to integrate edge/fog and cloud infrastructures to support the execution of heterogeneous IoT applications. 
\item Containerization-support for software components of the framework and IoT applications is proposed for fast deployment and efficient reuse of resources.
\item Dynamic scheduling, scalability, and resource discovery mechanisms are developed for fast adaptation as the characteristics of environment change.
\item A real-world prototype is developed using FogBus2 with a real-time IoT application named Conway's Game of Life, and a non-real-time IoT application, called Video Optical Character Recognition (VOCR). 
\end{itemize}
\par

\section{Evaluation Methodologies}
\label{sec:evaluattionMethodologies}
To conduct the experiments and evaluate the performance of our proposed framework, we had two potential approaches, namely simulation and real-world system setup. We first run all the virtual environment components, allocating different resources to virtual machines, simulating the heterogeneity of the integration edge/fog and cloud infrastructures. However, there was a considerable difference from the result of real-world infrastructures. Thus, after tried simulation and got no ideal results, we shifted our experimental setup to real-world devices, including Raspberry Pis, desktop, and cloud instances provided by different providers worldwide. The specifications of the infrastructures for experiments will be introduced in the performance evaluation sections, respectively.

The features of FogBus2 are evaluated using one real-world application we developed, named Conway's Game of Life, which is both computation-intensive and latency-sensitive. The application is implemented with 62 modules (tasks). For each data frame, every two modules execute for the computation of grids at the same size. The execution of modules with different sizes depending on their parent module's execution result. This structure strongly increases the complexity of the problem to schedule the execution of the application. 

Theoretically, when there are \textit{N} hosts distributed in the system, $N^{62}$ different potential solutions exist to execute Conway's Game of Life. Because the resource of each host and bandwidth/data rate between any two pair of hosts are not the same, it is impossible to find a good enough solution in a short time with a native approach such as brute force. Thus, we implemented two popular policies and compared them with our proposed policy to evaluate our framework. We use response time, a lower better metric, to assess the performance of scheduling. The details of this experiment will be presented in Chapter~\ref{sec:analysisOfSchedulingPolicies}.

However, the scheduling mechanism is not the only requirement of the framework to use edge/fog servers and cloud servers simultaneously efficiently. When there are numerous requests for the execution in the distributed system, the requests can not be put in a waiting queue too long, which brings users horrible experiences and wastes edge/fog servers and cloud servers by leaving them idle. Thus, the scalability mechanism is evaluated following the scheduling mechanism. Increasing numbers of simultaneous requests are tested in the experiment to determine how efficiently the scalability mechanism of FogBus2 is. We define and use Scheduling Finish Time (SFT). This metric is also lower better. The details will be presented in Chapter~\ref{sec:analysisOfMasterScalability}.

Moreover, as we use the containerization technique in the implementation, we experiment to evaluate how much Resource Ready Time (RRT) can be reduced when the containers for execution can be reused. In the experiment, we tested five different applications. Before requesting each of them, there are already execution containers in the cool-off period and ready to be reused. Thus, to verify the performance of the resources reuse mechanism, an experiment has been conducted to compare how long it would take for the resources placed to be ready when heterogeneous IoT applications are requested. With the defined metric, resource ready time, reuse mechanism is applied and bypassed for two IoT applications. The applications have different complexity regarding the number of modules (containers) required. The performance of our reuse mechanism will be examined in the experiment. The details will be shown in Chapter~\ref{sec:taskExeutorReuse}

Finally, startup time and RAM usages are studied in Chapter~\ref{sec:startupTimeAndRAMUsageAnalysis} comparing with FogBus \cite{tuli2019fogbus} which shows the FogBus2 framework is lightweight.

\newpage
\section{Thesis Organization}
\begin{figure}[t]
    \centering 
    \includegraphics[width=\linewidth]{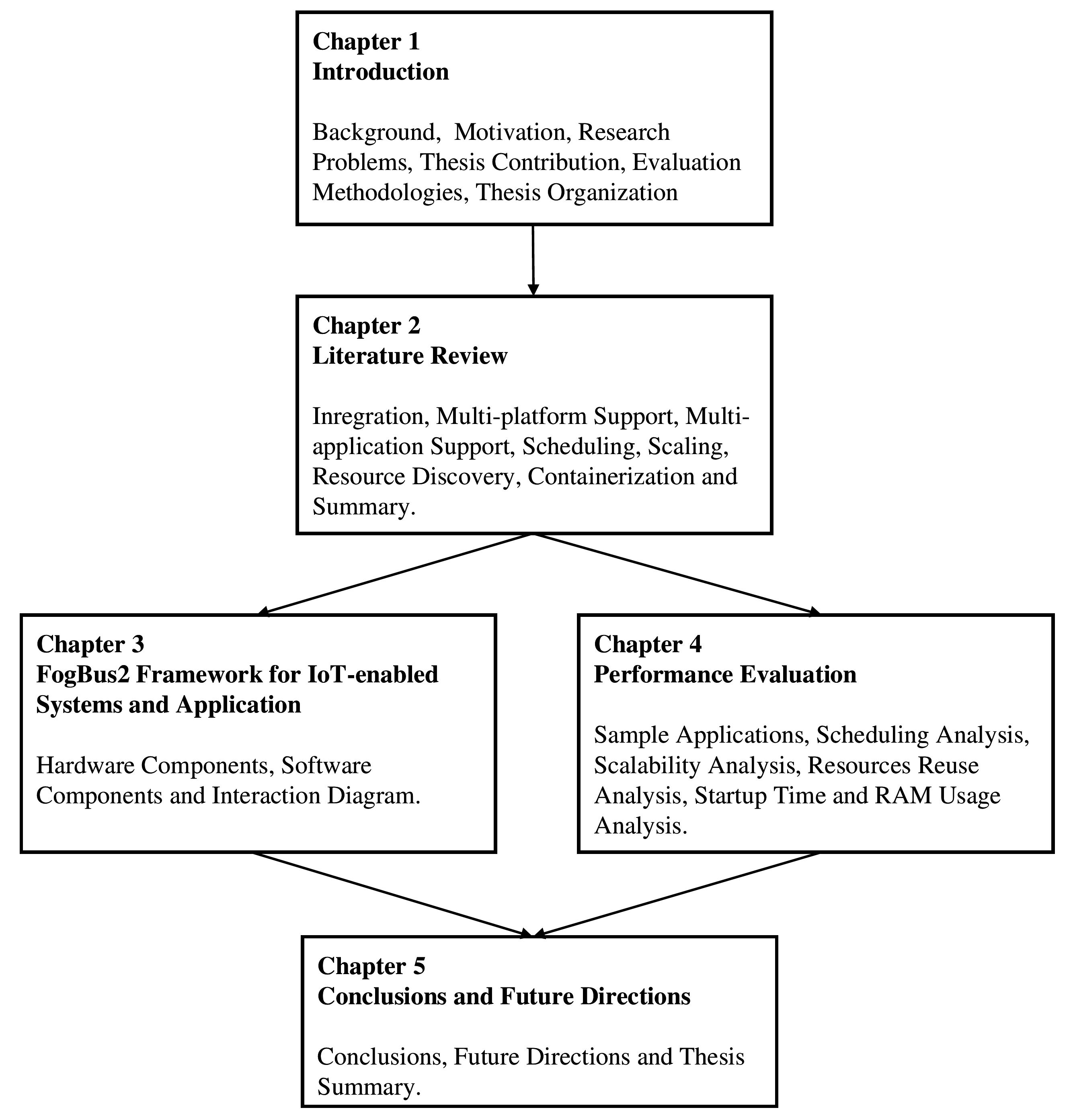}
    \caption{Visualized Thesis Organization}
    \label{fig:thesisStructure}
\end{figure}

The rest of the paper is organized as follows. Relevant frameworks are reviewed, and features of them are summarized in topologies in Chapter~\ref{chapt:relatedWork}. Chapter~\ref{chapt:FogBus2Framework} presents the hardware and software components of the FogBus2 and some of the details of framework design. The performance of FogBus2 is evaluated in Chapter~\ref{chapt:PerformanceEVAL}. Finally, Chapter~\ref{chapt:Conclusion} concludes the paper and draws future works. A visualized thesis organization is presented in Figure~\ref{fig:thesisStructure}.

\chapter{Literature Review}
\label{chapt:relatedWork}
This chapter discusses related frameworks integrating IoT-enabled systems with edge/fog and cloud infrastructures. We have discussed research questions that include 1) how to use edge/fog devices and cloud servers simultaneously efficiently, 2) how to provide a scheduling mechanism for incoming requests from different types of IoT applications, 3) how to provide a scalable platform in these heterogeneous computing environments, and 4) how to support automatically discover resources and reuse for such platform. To answer these questions, five aspects are identified as follows based on the importance to consider the heterogeneity of infrastructures, heterogeneity of IoT applications, efficient use of heterogeneous resources, intelligently discovering resources, and making the strength of containerization technologies:
\begin{itemize}
 \item \textbf{Integration} IoT applications produce a massive amount of data for processing and storage when edge computing exists extremely close to IoT devices. Edge/fog devices compute, store and form networks to serve IoT devices with low latency, high bandwidth, and distributed mobility. Cloud computing that provides unlimited computing and storage resources is necessary for computation-intensive applications with high performance even though cloud computing has low bandwidth and high latency. Seamless integration of edge/fog and cloud infrastructures to support different IoT applications is an important research topic. It is a challenge that a framework makes the most of strength by using the three efficiently in the framework design. Chapter~\ref{sec:integration} reviews frameworks that integrate IoT, edge/fog, and or the cloud.
 \item \textbf{Multi-platform Support} There are vast amounts of IoT devices, and they are gracefully embracing the heterogeneity of multiple platforms. The emergence of IoT devices brings a growing capacity of programming for distinct and different devices when they use various CPU architectures, and a wide variety of operating systems \cite{auler2017handling}. A framework designed for running on IoT devices, ECs, and cloud servers with high diversity needs to support multiple platforms to integrate edge/fog and cloud infrastructures to support the execution of heterogeneous IoT applications and helps to reduce the burden of developers to support a variety of platforms. Chapter~\ref{sec:multiPlatformSupport} introduces related frameworks which support multi-platform.
 \item \textbf{Multi-application Support} IoT devices diverse when they have various hardware and sensors when the data they produce and traffic demand also require different resources to transport and compute \cite{zhao2018deploying}. Consequently, IoT applications developed on top of various sensors and hardware are varied in a large number. A framework needs to support multiple IoT applications, which may be latency-sensitive, computation-intensive, and or bandwidth-sensitive. When different applications usually run simultaneously and distribute over the network, they can run separately or co-operate with each other. Thus, a framework needs to have the ability to support such different kinds of applications without requiring much effort that developers need to make. Chapter~\ref{sec:multiApplicationSupport} reviews related frameworks which support multi-application.
 \item \textbf{Scheduling Mechanism} Resources of distributed edge/fog devices and cloud servers are highly heterogeneous in computing capabilities, processors' architectures, RAM capacity, and supported communication protocols. The scheduling task is usually a complex undertaking to manage such countless and growing resources \cite{buyya2002gridsim}. Thus, an efficient scheduling mechanism is required to arrange heterogeneous IoT applications' execution using distributed edge/fog devices and cloud server resources. A scheduling algorithm (policy) real-timely configures the executions of different modules of an application, usually decides how to arrange and control the executions based on the current environment of the system, such as how many resources are available. The framework needs to obtain the ability to integrate multiple scheduling policies. Chapter~\ref{sec:schedulingMechanism} surveys related frameworks that use scheduling mechanisms.
 \item \textbf{Scaling Mechanism} The scaling mechanism is eminent to automatically keep Quality of Services (QoS) at a reasonable level which gives users a seamless experience when the workloads come into a system are unpredictable \cite{buyya2010intercloud}. The scaling mechanism in a framework is responsible for automatically adding or allocating resources when the workload feeding into a system is growing. IoT applications are close to the natural environment and humans' living environment. Therefore, the demand for the execution of the applications changes dynamically and unpredictably. A framework should have scaling mechanism support to dynamically control the distributed resources with such changing IoT applications execution demand. Chapter~\ref{sec:scalingMechanism} studies a framework on its scaling mechanism.
 \item \textbf{Resource Discovery Mechanism} Resource Discovery, sometimes called Service Discovery, is the capability to automatically find and locate resources and or services that are not predefined or addressed in a network \cite{chakrabarti1999focused}. IoT devices, edge/fog devices, and cloud servers are fully distributed geographically and logically; thus, the resource discovery mechanism empowers the hosts to discover others, form a subnet, and share their resources. A framework needs to obtain the capability of resource discovery mechanism trying to use distributed resources as efficiently as possible. Chapter~\ref{sec:resourceDiscovery} investigates a framework on its resource discovery mechanism.

 \item \textbf{Containerization Technique} Containerization is the technique that provides operating system-level isolation of a process, a program, or a complete operating system environment. It is as secure as a virtual machine but more lightweight since it does not depend on the host hardware emulation \cite{dua2014virtualization}. The containerization technique allows both the framework components and IoT applications to be containerized, which benefits faster deployments of IoT applications, easy configuration of the framework itself, and automatic scaling mechanism. A framework supporting the containerization technique can be quickly released when retaining and taking advantage of the features of the containerization technique. Chapter~\ref{sec:containerizationTechnique} statisticizes related frameworks which adopts containerization technique.
\end{itemize}

\section{Integration}
\label{sec:integration}
Tuli et al. proposed the FogBus framework based on a master-worker approach \cite{tuli2019fogbus}. This framework integrates IoT systems to the cloud and edge infrastructures when harnessing IoT resources, edge resources, and cloud resources. FogPlan, developed by Yousefpour et al., is a container-enabled framework integrating IoT devices with edge/fog devices and cloud devices to minimize the response time of IoT applications \cite{yousefpour2019fogplan}. Merlino et al. developed a framework recognizing offloading patterns \cite{merlino2019enabling}. They use a middleware platform to integrates IoT, edge/fog devices, and cloud devices, which tries to improve OpenStack and Stack4Things. Nguyen et al. \cite{nguyen2019market} proposed a privacy-preserving framework that handles requests and data in edge/fog devices and cloud devices, combining local execution at the edge and remote processing at the cloud. An et al. developed the EiF framework to bring artificial intelligence services to the edge of the network \cite{an2019eif}. It manages service dependencies and relations of IoT applications over the network of edge devices and cloud devices. Borthakur et al. developed the SmartFog framework, integrating IoT devices with edge/fog devices to analyze pathological speech data obtained from wearable sensors \cite{borthakur2017smart}. Yigitoglu et al. developed a container-enabled Foggy framework that provides automatic resource control in heterogeneous infrastructures such as IoT devices, edge/fog devices, and cloud devices \cite{yigitoglu2017foggy}. Bellavista et al. proposed a centralized framework extending Kura framework, and the design includes components running on IoT, edge/fog devices, and cloud devices \cite{bellavista2017feasibility}. Ferrer et al. developed an Adhoc-based framework to support the integration of IoT devices with multi-hop edge/fog devices \cite{ferrer2019ad}. Noor et al. developed a centralized container-enabled IoTDoc framework to manage interactions between IoT devices and cloud resources \cite{noor2019iotdoc}.

\section{Multi-platform Support}
\label{sec:multiPlatformSupport}
Tuli et al. proposed FogBus, which overcomes the difficulty of executing applications in heterogeneous infrastructures by developing the framework in a cross-platform programming language \cite{tuli2019fogbus}. By doing so, the framework can be barrier-freely deployed on heterogeneous infrastructures without any implementation changes. FogPlan pulls services (applications) which were automatically pushed to the cloud \cite{yousefpour2019fogplan}. This design ensures the new versions of their services are always fresh in the cloud database. This deployment mechanism masks the heterogeneity of platforms and enables the migration of their services between hosts. Merlino et al. proposed a middleware platform based on OpenStack, which encompasses edge/fog devices and cloud devices \cite{merlino2019enabling}. It is designed for big data processing in a hierarchical design involved cloud devices. In the design of the Foggy framework, the authors present a list of deployment workflow with the assumption that the required environment of applications is maintained by other Orchestration Server \cite{yigitoglu2017foggy}. By following the workflow, the framework helps developers reduce the effort to considering the complex heterogeneity of IoT infrastructures, edge/fog infrastructures, and cloud infrastructures.

\section{Multi-application Support}
\label{sec:multiApplicationSupport}
Nguyen et al. proposed a framework in which the edge/fog devices support content, services, and applications from different providers to serve their customers by proactively distribute the data into edge/fog devices \cite{nguyen2019market}. Bellavista et al. proposed a framework that adds brokers at gateways side trying to solve the problem that a flat topology is insufficient for both envisioned IoT applications and various real-word applications domain \cite{bellavista2017feasibility}. 

\section{Scheduling Mechanism}
\label{sec:schedulingMechanism}
Ferrer et al. designed a mechanism to utilize the existing capacity of edge/fog devices, which responses to the increasing demands from IoT with a massive volume of data \cite{ferrer2019ad}. FogPlan has the scheduling mechanism to deploy or release services by monitoring the incoming traffic and other parameters with the hypothesis that their service controllers only maintain the edge/fog hosts in a particular topographical subnet\cite{yousefpour2019fogplan}. Nguyen et al. developed a centralized resource allocation technique that considers the current resources of edge/fog devices and cloud devices\cite{nguyen2019market}. An et al. proposed the SmartFog framework, which uses an unsupervised clustering method analyzing the lower-resources workload on Intel Edison and Raspberry Pi \cite{borthakur2017smart}. Foggy integrates a mechanism to scheduling tasks aiming to optimize overall resources utilization, minimize latency between IoT, edge/fog devices, and cloud devices \cite{yigitoglu2017foggy}.

\section{Scaling Mechanism}
\label{sec:scalingMechanism}
Bellavista et al. use docker containers and the Kubernetes to scale computing infrastructures that support geographically distributed IoT applications and their deployment mechanism \cite{bellavista2017feasibility}. FogPlan supports a simple mechanism of scalability based on the monitored aggregated traffic rate, which uses the ping approach \cite{yousefpour2019fogplan}. However, FogPlan does not support policy integration which makes the scaling mechanism fixed and not extensible.

\section{Resource Discovery}
\label{sec:resourceDiscovery}
The middleware discovers resources when assumes edge/fog devices join or leave the system unpredictably \cite{merlino2019enabling}. Using a Cloud Manager monitoring subsystem, the discovery is triggered when a signal shows the reducing synthetic manipulation performance. FogPlan has mechanism of edge/fog resources discovery. The hosts at edge advertise their IP addresses to IoT devices that run IoT applications \cite{yousefpour2019fogplan}. FogPlan also carries URI in communication protocol routing the requests to edge hosts, which achieves resource discovery.

\section{Containerization Technique}
\label{sec:containerizationTechnique}
FogPlan has the assumption that the applications running on their framework are containerized, which allows to automate the deployment and to release of services \cite{yousefpour2019fogplan}. The services are also considered stateless; thus, migration, deployment, and release of the services are faster than VM-based migration procedures. Merlino et al. containerized edge/fog devices and cloud devices resources in their work \cite{merlino2019enabling}. Foggy uses containerization techniques to detach subnets and attempts to decrease the overhead on the constraint resource among hosts over edge/fog devices, and cloud devices \cite{yigitoglu2017foggy}. Docker containers and the Kubernetes technologies are used by Bellavista et al. to manage the execution of various IoT applications over the heterogeneous resources of edge/fog devices and cloud devices \cite{bellavista2017feasibility}. Ferrer et al. employ workload virtualization, which is containerized to facilitate the execution in heterogeneous edge/fog devices environments \cite{ferrer2019ad}. Noor et al. proposed a framework that is also developed with containerization technology \cite{noor2019iotdoc}.

\section{Summary}

Table~\ref{tab:relatedWork} identifies and compares the main elements of related frameworks with ours. These frameworks often do not support platform-independent techniques and or containerization of software components of the framework and IoT applications. Moreover, most of these frameworks do not offer scheduling, scalability, and resource discovery mechanisms. FogBus2 provides a lightweight and container-enabled distributed framework for computation-intensive and latency-sensitive IoT applications to overcome these limitations. It dynamically schedules heterogeneous IoT applications and scales the resources to serve IoT users efficiently.
\label{sec:litSummary}
\begin{center}
 \begin{table}
    \centering
        \caption{A qualitative comparison of related works with ours}
    \label{tab:relatedWork}

    \resizebox{\linewidth}{!}{%
        \begin{tabular}{|c|c|c|c|c|c|c|c|} 
            \hline
            Work             & Integration                                               & \begin{tabular}[c]{@{}c@{}}Multi\\ Platform \\ Support \end{tabular} & \begin{tabular}[c]{@{}c@{}}Heterogeneous \\ Multi\\ application\\ Support \end{tabular} & \begin{tabular}[c]{@{}c@{}}Dynamic \\ Scheduling\\ Mechanism and \\ Policy Integration \end{tabular} & \begin{tabular}[c]{@{}c@{}}Dynamic\\ Scaling \\ and Policy\\ Integration \end{tabular} & \begin{tabular}[c]{@{}c@{}}Dynamic\\ Resource\\ Discovery \end{tabular} & \begin{tabular}[c]{@{}c@{}}Container\\ Support \end{tabular}  \\ 
            \hline
            \cite{tuli2019fogbus}           & \begin{tabular}[c]{@{}c@{}}IoT, Edge,\\Cloud\end{tabular}   & \checkmark                                                              & $\times$                                                                                      & $\times$                                                                                                  & $\times$                                                                                      & $\times$                                                                        & $\times$                                                             \\ 
            \hline
            \cite{yousefpour2019fogplan}& \begin{tabular}[c]{@{}c@{}}IoT,  Edge,\\Cloud\end{tabular}   & \checkmark                                                                    & $\times$                                                                                       & \checkmark                                                                                                  & $\times$                                                                                    & \checkmark                                                                        & \checkmark                                                             \\
            \hline
            \cite{merlino2019enabling} & \begin{tabular}[c]{@{}c@{}}IoT,  Edge,\\Cloud \end{tabular} & \checkmark                                                                    & \checkmark                                                                                       & $\times$                                                                                                  & $\times$                                                                                      & \checkmark                                                                        & \checkmark                                                             \\ 
            \hline
            \cite{nguyen2019market} & \begin{tabular}[c]{@{}c@{}}IoT,  Edge,\\Cloud \end{tabular} & \checkmark                                                                    & \checkmark                                                                                      & \checkmark                                                                                                  & $\times$                                                                                      & $\times$                                                                        & $\times$                                                             \\ 
            \hline
            \cite{an2019eif} & \begin{tabular}[c]{@{}c@{}}IoT,  Edge,\\Cloud \end{tabular} & $\times$                                                                    & $\times$                                                                                       & $\times$                                                                                                  & $\times$                                                                                      & $\times$                                                                        & $\times$                                                             \\ 
            \hline
            \cite{ghosh2019mobi} & \begin{tabular}[c]{@{}c@{}}IoT,  Edge,\\Cloud \end{tabular} & $\times$                                                                   & $\times$                                                                                       & \checkmark                                                                                 & $\times$                                                                                      & $\times$                                                                        & $\times$                                                             \\ 
            \hline
            \cite{borthakur2017smart} & IoT,  Edge                                                  & \checkmark                                                                 & $\times$                                                                                       & $\times$                                                                                                  & $\times$                                                                                      & $\times$                                                                        & $\times$                                                             \\ 
            \hline
            \cite{yigitoglu2017foggy} & \begin{tabular}[c]{@{}c@{}}IoT,  Edge,\\Cloud \end{tabular} & \checkmark                                                                    & $\times$                                                                                       & \checkmark                                                                                                  & $\times$                                                                                      & $\times$                                                                        & \checkmark                                                             \\ 
            \hline
            \cite{bellavista2017feasibility} & \begin{tabular}[c]{@{}c@{}}IoT,  Edge,\\Cloud \end{tabular} & $\times$                                                                   & \checkmark                                                                                       & $\times$                                                                                                 & $\checkmark$                                                                                      & $\times$                                                                        & \checkmark                                                             \\ 
            \hline
            \cite{ferrer2019ad} & IoT,  Edge                                                  & $\times$                                                                    & $\times$                                                                                       & \checkmark                                                                                                  & $\times$                                                                                      & $\times$                                                                        & \checkmark                                                             \\ 
            \hline
            \cite{noor2019iotdoc}& IoT, Cloud                                                & $\times$                                                                    & $\times$                                                                                       & \checkmark                                                                                                  & $\times$                                                                                      & $\times$                                                                        & \checkmark                                                             \\  
            \hline
            FogBus2          & \begin{tabular}[c]{@{}c@{}}IoT,  Edge,\\Cloud \end{tabular} & \checkmark                                                                   & \checkmark                                                                                       & \checkmark                                                                                                  & \checkmark                                                                                      & \checkmark                                                                        & \checkmark                                                             \\
            \hline
        \end{tabular}
    }
\end{table}

\end{center}
Tuli et al. proposed the FogBus framework based on a master-worker approach to process data generated from sensors on edge/fog devices or cloud devices \cite{tuli2019fogbus}. Due to platform-independent technologies used in the FogBus, it can work on multiple platforms. However, it does not provide any mechanism for dynamic scheduling of IoT applications, scalability, and resource discovery. Besides, it does not support different communication topologies between workers and the master. Moreover, FogBus is not a container-enabled framework, which negatively affects the deployment cost of IoT applications and software components. Yousefpour et al. developed a container-enabled framework, called FogPlan, integrating IoT devices with edge/fog devices and cloud devices to minimize the response time of IoT applications \cite{yousefpour2019fogplan}. FogPlan supports dynamic resource discovery, scheduling of IoT applications, and simple scalability mechanism and policies. Merlino et al. developed a container-enabled framework for container discovery at edge/fog devices and cloud devices, and horizontal and vertical offloading \cite{merlino2019enabling}. However, it does not provide any policies for the dynamic scheduling of IoT applications and the scalability of resources. Nguyen et al. proposed a privacy-preserving framework, which uses obfuscation to keep users' information private meanwhile tasks are computed \cite{nguyen2019market}. Besides, they developed a centralized resource allocation technique that considers the current resources of edge/fog devices and cloud devices. An et al. developed the EiF framework to bring artificial intelligence services to the edge of the network \cite{an2019eif}. Although the EiF provides some resource allocation techniques for network resources, it does not offer any scheduling and scalability mechanisms for IoT applications. A mobility-aware framework, called Mobi-IoST, is developed by Ghosh et al. \cite{ghosh2019mobi}, which uses a probabilistic approach for the placement of IoT applications. Borthakur et al. developed the SmartFog framework, integrating IoT devices with edge/fog devices to analyze pathological speech data obtained from wearable sensors \cite{borthakur2017smart}. It embeds machine learning techniques to analyze the generated data at the proximity of patients. Yigitoglu et al. developed a container-enabled Foggy framework that supports dynamic scheduling of containerized IoT applications with dependent tasks \cite{yigitoglu2017foggy}. Bellavista et al. proposed a centralized container-enabled framework that uses docker containers and the Kubernetes to scale computing infrastructures \cite{bellavista2017feasibility}. However, it does not provide any policies to support scalability, scheduling, and resource discovery.
Moreover, as the cloud orchestrator manages the deployments of applications, it may negatively affect the response time of latency-sensitive IoT applications. Ferrer et al. developed a container-enabled Adhoc-based framework to support the integration of IoT devices with multi-hop edge/fog devices \cite{ferrer2019ad}. Noor et al.developed a centralized container-enabled IoTDoc framework to manage interactions between IoT devices and cloud resources \cite{noor2019iotdoc}.

\chapter{FogBus2 Framework for IoT-enabled Systems and Applications}
\label{chapt:FogBus2Framework}

\begin{center}
    \begin{figure}[t]
	\centering 
	\includegraphics[width=\linewidth]{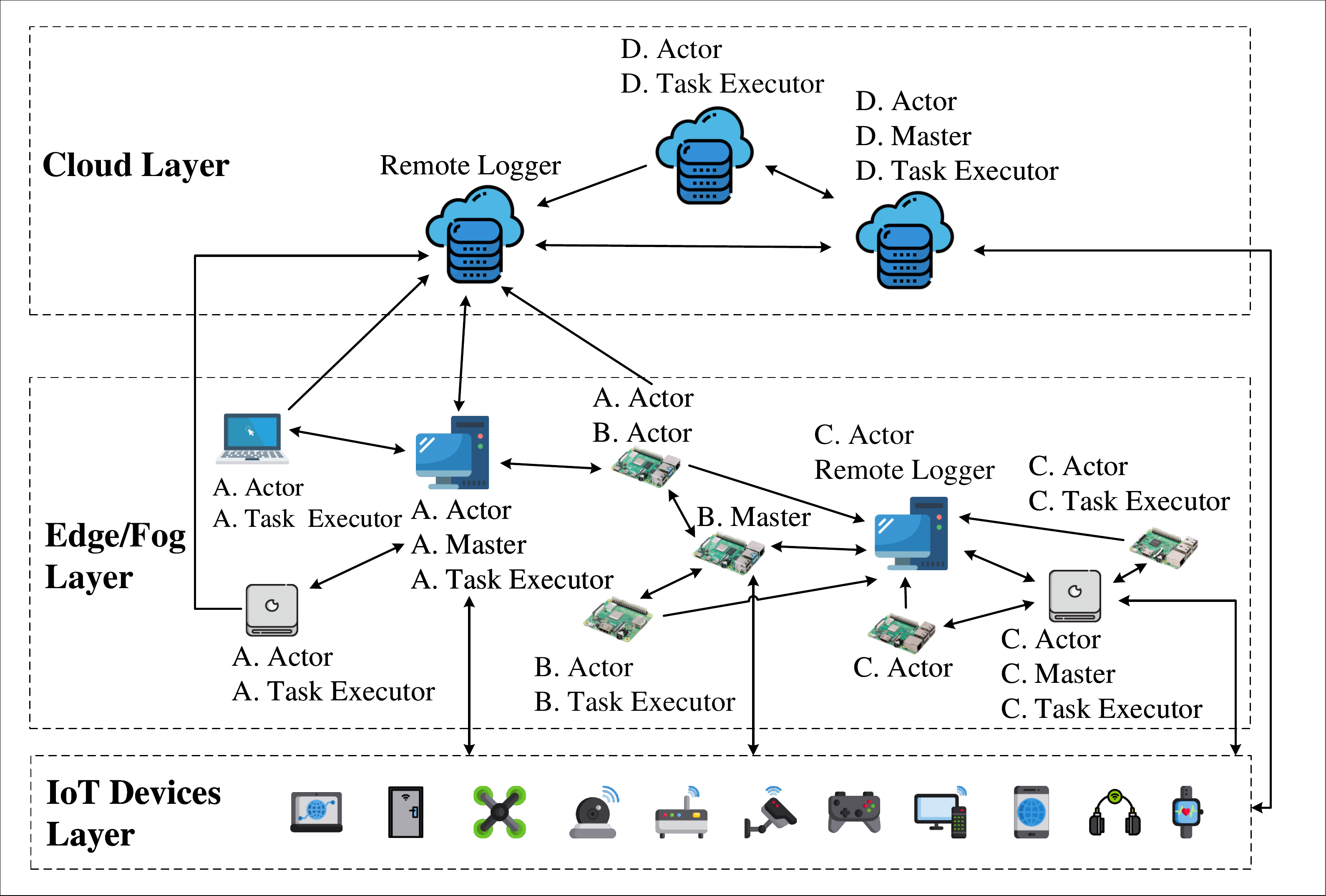}
	\caption{FogBus2 high-level computing environment}
	\label{fig:highLevelFramework}
\end{figure} 

\end{center}

This chapter describes the hardware and software components of FogBus2 in detail. Chapter~\ref{sec:hardware} presents the hardware components and related design of the framework. Fig.~\ref{fig:highLevelFramework} presents a high-level overview of computing environment supported by FogBus2. Chapter~\ref{sec:software} presents components and sub-components of the framework design.

\section{Hardware Components}
\label{sec:hardware}
FogBus2 supports heterogeneous hardware resources such as different IoT devices, Edge/Fog servers, and multiple cloud data centers. In Chapter~\ref{chapt:PerformanceEVAL}, devices of three of the infrastructures in the real world have been used to run FogBu2's components. Since the containerization technique is used in our framework for all the components, FogBus2 should also be compatible with other devices that are supported by the containerization technique. As Fig.~\ref{fig:highLevelFramework} shows, multiple different subnets are existing over the three layers simultaneously. And one particular host can run several components according to what resources it owns.

\subsection{IoT devices layer}
IoT devices layer consists of heterogeneous types of resource-limited IoT devices (such as drones, smart cars, smartphones, security cameras, any types of sensors such as humidity sensors, etc) that perceive data from the environment and perform physical actions on the environment. FogBus2 provides a distributed platform for IoT devices to connect with proximate and remote service providers through different communication protocols such as WiFi, Bluetooth, Zigbee, etc. Hence, the generated data from IoT devices can be processed and stored on surrogate servers with higher resources, which significantly helps to reduce the processing time of data generated from IoT devices. 

In this layer, the IoT devices play the role of $user$ when they sense the data from the internal or external environment and request various applications running on the system. After the requests have been approved and resources placed are ready, devices in this layer send data to the system and receive the result. The actuator of a device itself will finally consume the result, and or actuators in other peers will do.

\subsection{Edge/Fog layer}
FogBus2 provides IoT devices with low-latency and high-bandwidth access to heterogeneous edge/fog resources distributed in their proximity. These heterogeneous edge/fog servers can be either one-hop away from IoT devices (such as Raspberry pis (RPi), personal computers, etc.) or multi-hop away (such as routers, gateways, etc.). Moreover, to extend the computing and storage capacity of edge/fog servers, FogBus2 supports the collaborative execution of IoT applications among different edge/fog servers in a distributed manner. Hence, FogBus2 offers a wide range of service options for different types of IoT devices with heterogeneous service-level requirements. 

FogBus2 components are primarily running in this layer since they are close to IoT devices, with low-latency and high-bandwidth advantages. By collaborating with others hosts in this layer, the components running in this layer dramatically boost the execution of IoT applications.

\subsection{Cloud layer}
FogBus2 expands IoT devices' computing and storage resources by supporting multiple cloud data centers in different geo-location areas, which bring location-independency for IoT applications. Moreover, cloud resources can either be used to process and or store computation and or storage-intensive tasks or when the edge/fog servers resources become overloaded. When the requested application from $user$ components from IoT applications is computation-intensive, and there is a lack of resources, the execution of the application is likely to be arranged to the cloud layer.

\section{Software Components}
\label{sec:software}
\begin{center}
    \begin{figure}[t]
    \centering 
    \includegraphics[width=\linewidth]{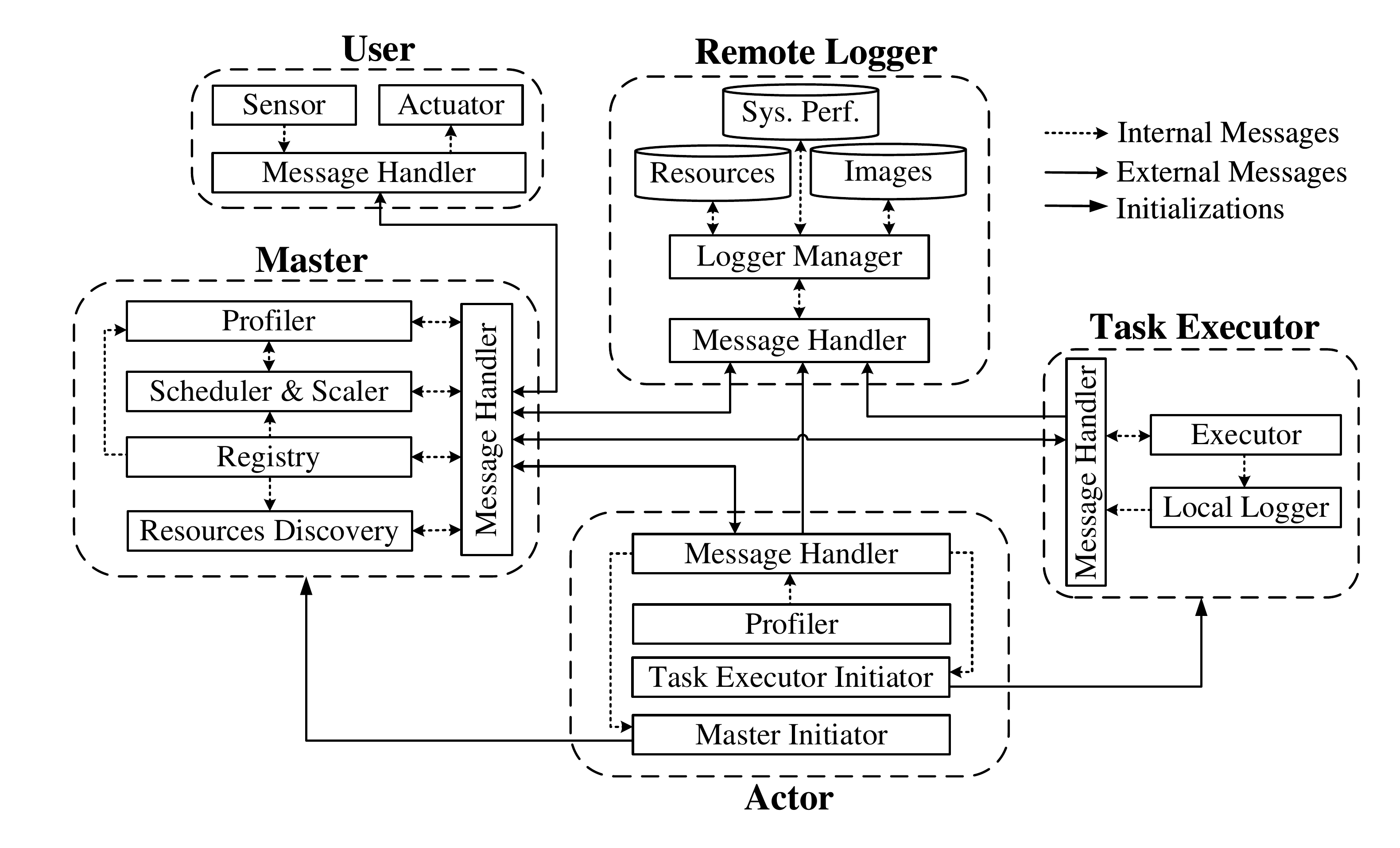}
    \caption{FogBus2 software components and interactions}
    \label{fig:LowLevelFramework}
\end{figure} 

\end{center}

FogBus2, which is developed from scratch, consists of five main containerized components (using the docker containers technique) developed in Python. Since FogBus2 is a distributed framework, these components can run on different hosts based on the application scenario, as depicted in Fig.~\ref{fig:highLevelFramework}. FogBus2's main components, sub-components (Sub-C), and their respective interactions is shown in Fig.~\ref{fig:LowLevelFramework}. In each component, a \textit{message handler} Sub-C is embedded for inter-component communications. 

\subsection{User component}
This component runs on $user$s' IoT devices and consists of \textit{sensor} and \textit{actuator}. It can send placement requests to the \textit{master} component for each IoT application, developed with either dependent or independent tasks. Also, it handles the sensors' raw data and collects the processed data from \textit{master}.  
\subsubsection{Sensor}
This Sub-C controls the sensing intervals of physical sensors and captures and serializes the sensors' raw data.

\subsubsection{Actuator} 
This Sub-C collects processed data from \textit{master} and executes an action based on the application scenario. To support multiple application scenarios, the \textit{actuator} can perform actions in real-time or perform periodic actions based on aggregated data.  

\subsection{Master Component}
This component can run on any hosts, either in edge/fog or cloud layers, based on the application scenario. It dynamically profiles the environment and performs resource discovery to find available computing and storage resources. Besides, the \textit{master} component receives placement requests from IoT devices, schedules them, and manages the execution of IoT applications.   

\subsubsection{Registry}
When the \textit{master} receives joining requests from \textit{actors} or \textit{task executors}, it records their information and assigns them a unique identifier for the rest of communications. Moreover, it handles placement requests of \textit{users}, assigns them a unique identifier, and initiates the \textit{$scheduler\;\&\; scaler$}. The \textit{master} uses each \textit{user'}s unique identifier to distinguish heterogeneous data arriving from other \textit{users}. Also, it can manage authentication mechanisms for the \textit{actors} and \textit{task executors}.

\subsubsection{Profiler}
This Sub-C initially receives information about available resources (such as CPU specifications, RAM), network characteristics (such as average bandwidth and latency), and IoT applications' properties (such as the number of tasks, dependency models) from \textit{registry} Sub-C. Afterward, the \textit{profiler} periodically updates its information from stored data in the \textit{remote logger} component. Moreover, if the required data is not available in the \textit{remote logger} or the \textit{master} requires updated information, it can directly communicate with IoT devices, \textit{actors}, or \textit{task executors} to obtain the data. Also, it keeps track of the status of the \textit{master} and its available resources.

\subsubsection{Scheduler}
When the IoT \textit{user} registered in the \textit{master}, its placement request will be forwarded to the $scheduler\;\&\; scaler$ and will be queued based on First-In-First-Out (FIFO) policy. Algorithm~\ref{alg:scheduler} 
describes the scheduling mechanism and the integrated Optimized History-based Non-dominated Sorting Genetic Algorithm (\textit{OHNSGA}) scheduling policy. The \textit{scheduler} de-queues each placement request based on the FIFO policy. Next, the \textit{scheduler} receives the list of \textit{actors} from the \textit{registry} Sub-C, and continues the scheduling procedure if there exists at least one registered \textit{actor}. Otherwise, it notifies the \textit{user} that there are not enough resources for the scheduling (lines 1-4). Afterward, the \textit{scheduler} examines the local resources of the host. If the CPU utilization is above the threshold ($max\_cpu\_util$) or the received placement requests exceeds the threshold ($max\_shed\_count$), it attempts to find a substitute \textit{master} ($sub\_master$) to serve this request in order to reduce the waiting time of \textit{user}'s placement request in the queue. If there exists other \textit{master} components in the computing environment, it attempts to find the best $sub\_master$ (with lowest access latency), otherwise it runs the \textit{scaler} to initiate a new \textit{master} component. (lines 5-12). If the current host has enough resources for the scheduling, the \textit{scheduler} retrieves the application and its dependency model (for IoT applications with dependent tasks) from the placement request. Moreover, it finds the list of \textit{actors} that can serve each task of an IoT application and stores them in $task\_actrs\_map$ (lines 13-21). The \textit{scheduler} then retrieves the history of previous decisions for this application (line 22). Next, the \textit{scheduler} initiates the \textit{OHNSGA} to find a suitable set of \textit{actors} for the IoT application to minimize its response time. (line 23). The response time of an IoT application is defined as the time difference when a \textit{user} component starts sending data to the time it receives the result.

 \begin{center}
    \begin{algorithm}[t]
    \footnotesize
    \caption{Scheduler} \label{alg:scheduler}
    \SetKwData{Left}{left}
    \SetKwData{This}{this}
    \SetKwData{Up}{up}
    \SetKwFunction{Union}{Union}
    \SetKwFunction{FindCompress}{FindCompress}
    \SetKwInOut{Input}{Input}
    \SetKwInOut{Output}{Output}
    \SetKwInOut{Parameter}{Parameter}
    \tcc{
        $req$: user request, 
        $prev\_dec$: decisions history, 
        $prof$: hosts profiles, 
        $curr\_sched\_count$: current scheduling threads count, 
        $max\_sched\_count$: max scheduling threads count, 
        $curr\_cpu\_util$: current CPU utilization, 
        $max\_cpu\_util$: max CPU utilization,
        $dependencies$: tasks dependencies,  
        $task\_actrs\_map$: map task to actors
    } 
        
        $actrs \gets $ \textsc{GetAllActors()} \\
        \If{$actrs$ is empty}{	
            \textsc{WarnUser}($req$) \\
            \Return \\
        }
        
        $curr\_cpu\_util\gets$ \textsc{GetCPUUtilization()} \\
        $curr\_sched\_count\gets$ \textsc{GetScheduleCount()} \\

        \tcc{If busy, forward request or scale a new Master}
        \If{ $curr\_cpu\_util > max\_cpu\_util$ or $curr\_sched\_count > max\_sched\_count$}{
                $sub\_master \gets$ \textsc{GetBestMaster($req, masters$)} \\
            \If{$sub\_master$ is null}{
                $sub\_master \gets$ \textsc{Scaler($req, actrs$)} \\
            }
            \textsc{NotifyUser($req, sub\_master$)} \\
            \Return \\
        }
        
        \tcc{Otherwise  schedule} 
        $dependencies, task\_list \gets$  \textsc{GetDependenciesAndTaskList($req$)} \\
        $i, task\_actrs\_map \gets  0, []$ \\
        \ForEach{$task\_list$}{
            $j, task\_actrs\_map[i] \gets 0, []$ \\
            \ForEach{$actrs$}{
                \If{$actr$ has image of $task$}{
                    $task\_actrs\_map[i][j] \gets actr$ \\
                    $j \gets j + 1$ \\
                }
            }
            $i \gets  i + 1$ \\
        }
        
        \tcc{Use OHNSGA to schedule} 
        $prev\_dec\gets$  \textsc{LoadHistory($req$)}   \\
        $res\gets$  \textsc{OHNSGA($prev\_dec, pop\_size, prof, task\_actrs\_map,  req$ )}  \\

        \For{$k$ from $0$ to $i-1$}{
            $actr \gets res[k]$ \\
            $task\_exec\_list \gets$ \textsc{GetIdleList($actr, task\_list[k]$)} \\
            
            \If{$task\_exec\_list$ is empty}{
                \textsc{SendInitTaskExecutorMsg($actr, task\_list[k], dependencies$)} \\
                continue
            }
            \textsc{SendReuseTaskExecutorMsg($task\_exec\_list[0],  dependencies$)} \\
        }
\end{algorithm}

\end{center}

 The \textit{OHNSGA} works based on a genetic algorithm (GA) which is a population-based evolutionary algorithm. Each candidate solution for assignments of \textit{actors} to tasks is called an individual, and the set of candidate individuals creates the population. The \textit{OHNSGA} attempts to find better individuals in each iteration of the algorithm to converge to the best solution. \textit{OHNSGA} uses the history of previous decisions of each application to initialize a portion of the first population while the rest of the population is randomly generated. It helps the \textit{OHNSGA} to start from a better initial state and reduces the convergence time of this technique. Also, as a portion of the population is randomly generated, the \textit{OHNSGA} keeps the randomness as well, which significantly helps to jump out of local-optimal solutions. The \textit{OHNSGA} uses the Tournament selection method to find the best individuals in each iteration. Then, to generate the population of the next iteration, \textit{OHNSGA} uses the Simulated Binary Crossover operator, that its efficiency is proved in \cite{deb2013evolutionary}, and Polynomial mutation operator. Algorithm~\ref{alg:ohnsga} presents an overview of the \textit{OHNSGA}. According to the outcome of \textit{OHNSGA}, the scheduler notifies the actors to run task executors or reuse the available ones for the current IoT application (lines 24-30).  

\begin{center}
    \begin{algorithm}[!t]
    \footnotesize
    \caption{OHNSGA} \label{alg:ohnsga}
    \SetKwData{Left}{left}
    \SetKwData{This}{this}
    \SetKwData{Up}{up}
    \SetKwFunction{Union}{Union}
    \SetKwFunction{FindCompress}{FindCompress}
    \SetKwInOut{Input}{Input}
    \SetKwInOut{Output}{Output}
    \SetKwInOut{Parameter}{Parameter}
    \tcc{
        $hist\_ratio$: ratio indicating the number of individuals generated based on history, 
        $init\_pop$: initial population, 
        $n\_offsprings$: number of offsprings,
        $pop$: population
    }
    $max\_num \_hist\_indv\gets \ceil[\big]{pop\_size / hist\_ratio} $ \\
    \If{$len(prev\_dec)  > max\_num \_hist\_indv $}{
        $prev\_dec \gets prev\_dec[0:max\_num \_hist\_indv]$	\\
    }
    
    $random\_indv \gets$ \textsc{RandomIndiv($pop\_size -  len(prev\_dec) $)} \\
    $init\_pop \gets $ \textsc{Merge($prev\_dec, random\_indv$)} \\
    $pop \gets $ \textsc{RemoveDuplicates($init\_pop$)}  \\
    \For{$i$ from 0 to $max\_iteration\_num$}{
        \While{True}{
            $parents \gets $ \textsc{TournamentSelection($pop, n\_parents$)} \\
            $offsprings \gets $ \textsc{SimBinCrossover($parents, n\_offsprings$)} \\
            $offsprings \gets $ \textsc{PolynomialMutation($offsprings$)} \\
            $pop \gets $ \textsc{Merge($parents, offsprings$)}  \\
            $pop \gets$ \textsc{RemoveDuplicates($pop$)}  \\
            \If{$len(pop) >= pop\_size$}{
                $pop \gets pop\_size[0:pop\_size]$  \\
                break \\
            }
        }
    }
    $pop \gets \textsc{Sort($pop$)}$ \\
    \Return $pop[0]$ 
    
\end{algorithm}

\end{center}
\subsubsection{Scaler}

The scaler is called when the current master requires to initiate a new master container. Algorithm~\ref{alg:scaler} depicts how scaler works. The scaler receives the list of registered actors and iterates over them to find the actor with the minimum latency and highest score. The scaler first considers the access latency of actors (line 7). Then, if the latency of the actor is equal to or less than the best-obtained latency, the scaler calculates a score value for that actor. The score value is obtained from current CPU utilization and the average CPU frequency of the host on which the actor is running (lines 8-12). Finally, the scaler selects the actor with the minimum latency whose score is higher and sends a message to the chosen actor to initiate a master container. 

\subsubsection{Resource Discovery}
\vspace{-1.5cm}
\begin{center}
    \begin{algorithm}[!t]

    \footnotesize
    \caption{Scaler} \label{alg:scaler}
    \SetKwData{Left}{left}
    \SetKwData{This}{this}
    \SetKwData{Up}{up}
    \SetKwFunction{Union}{Union}
    \SetKwFunction{FindCompress}{FindCompress}
    \SetKwInOut{Input}{Input}
    \SetKwInOut{Output}{Output}
    \SetKwInOut{Parameter}{Parameter}
    \tcc{
        $my\_addr$: address of this host, 
        $cpu\_util$: CPU utilization, 
        $cpu\_freq$: CPU frequency
    }
    $best\_actr\gets actrs[0]$\\ 
    $cpu\_util \gets$ \textsc{GetCPUUtilization($best\_actr$)}\\
    $cpu\_freq \gets $ \textsc{GetCPUFrequency($best\_actr$)}\\
    $best\_score \gets (1-cpu\_util) * cpu\_freq$ \\
    
    $min\_latency \gets$ \textsc{FindLatency($user$, $best\_actr$)} \\
    \ForEach{$actrs$}{
        $latency \gets$ \textsc{FindLatency($actr$)} \\
        \If{$latency > min\_latency$}{
            continue
        }
                        $cpu\_util\gets$ \textsc{GetCPUUtilization($actr$)}\\
            $cpu\_freq\gets$ \textsc{GetCPUFrequency($actr$)}\\
            $score =(1- cpu\_util) * cpu\_freq$ \\
        \If{$latency == min\_latency$ and $score<  best\_score$}{
                continue
            }
        
        $best\_actr \gets  actr$ \\
        $best\_score \gets score$\\
        $min\_latency \gets latency$ \\
    }

    \textsc{SendInitNewMasterMsg($best\_actr, my\_addr$)} 
    
\end{algorithm}

\end{center}
The key responsibility of this Sub-C is to find \textit{master} and \textit{actor} containers in the network. Algorithm~\ref{alg:resourceDiscovery} describes how resource discovery periodically works. This Sub-C receives the list of its registered \textit{actors} from the \textit{registry} (line 8). Then, it examines the network to find the list of all available neighbors (line 9). Next, this Sub-C checks each neighbor to find running \textit{master} and \textit{actor} containers. If the neighbor runs the \textit{master} container, the resource discovery adds the neighbor to its $known\_masters$ list and receives the list of registered \textit{actors} on the neighbor (lines 12-14). This mechanism helps \textit{master} containers to automatically know each other in the network and share the information of their registered \textit{actors}. Besides, if the neighbor runs the \textit{actor} container, the address of the \textit{actor} will be recorded in $new\_actrs$. Finally, the resource discovery Sub-C advertises the \textit{master} to all \textit{actors} that are not registered in its \textit{actor} list, $actrs$ (lines 17-19). 

Resource Discovery discovers a component by sending a probing message to an address. Any running component responds with its component information to the source address where the probing message is sent from. With this mechanism, resource discovery scans the network periodically and thus obtain the ability to discover new joined resources.

\begin{center}
    \begin{algorithm}[!t]
    \footnotesize
    \caption{Resource Discovery} \label{alg:resourceDiscovery}
    \SetKwData{Left}{left}
    \SetKwData{This}{this}
    \SetKwData{Up}{up}
    \SetKwFunction{Union}{Union}
    \SetKwFunction{FindCompress}{FindCompress}
    \SetKwInOut{Input}{Input}
    \SetKwInOut{Output}{Output}
    \SetKwInOut{Parameter}{Parameter}
    \algblock[TryCatchFinally]{try}{finally}
    \algcblockdefx[TryCatchFinally]{TryCatchFinally}{catch}{finally}
    [1]{\textbf{catch}}{}
    \tcc{
        $prev\_ad\_ts$: timestamp of the previous advertising, 
        $actrs$: all registered actors in current master,
        $neighbours$: neighbours in the network, 
        $interval$: discovery period
    }

    $prev\_ad\_ts \gets$  \textsc{Timestamp()}\\

    \While{True}{
        \tcc{Sleep for an interval}
        $ts\gets\textsc{Timestamp()}$\\ 
        \If{$ts - prev\_ad\_ts < interval$}{
            \textsc{SleepForAWhile()} \\ 
            continue \\
        } 
        \tcc{Record current timestamp}
        $prev\_ad\_ts \gets ts$  \\
        $actrs \gets$ \textsc{GetAllActors()} \\
        \tcc{Advertise itself to neighbours}
         $neighbours \gets $ \textsc{GetAllHosts}($net\_gateway$, $net\_mask$)  \\
        $new\_actrs \gets []$\\
        \ForEach {$neighbours$}{
            \If{$neighbour$ is Master}{
                $known\_masters \overset{+}{\leftarrow}$ $neighbour$ \\
                $new\_actrs \overset{+}{\leftarrow}$  \textsc{GetActorsAddrFrom($neighbour$)} \\
            } 
            
            \If{$neighbour$ is Actor}{
               $new\_actrs \overset{+}{\leftarrow}$  $neighbour$ \\
            }

            }
        \ForEach{$new\_actrs$}{
            \If{$new\_actr$ is not in $actrs$}{
                \textsc{AdvertiseSelf($new\_actr$ )}
            } 
            
        }
    }  
\end{algorithm}

\end{center}

\subsection{Actor component}
This component can run on any hosts in edge/fog or cloud layers. The \textit{actor} profiles the host's resources and starts the \textit{task executors} for the execution of IoT applications' tasks. Besides, it can initiate the \textit{master} container on the host for the scalability scenarios.

\subsubsection{Profiler}
This \textit{actor's profiler} works the same as the \textit{master's profiler} and records the available resources of the host and network characteristics. However, contrary to \textit{master's profiler}, it does not have profiling information of other hosts. The \textit{actor} periodically sends its profiling information to the \textit{remote logger} component as well as $master$ component where it has registered at.  

\subsubsection{Task executor initiator}
Whenever a \textit{master} component assigns a task of an IoT application to an \textit{actor} for the execution, the \textit{task executor initiator} is called. It initiates the \textit{task executor} and defines where the results of the \textit{task executor} should be forwarded.

\subsubsection{Master initiator}
This Sub-C is only called when a \textit{master} component (e.g., \textit{master A}) runs its \textit{scaler} procedure and decides to initiate a new \textit{master} component (e.g., \textit{master B}) on other hosts. Hence, the selected \textit{actor} receives a message from its \textit{master} component (\textit{master A}) and runs \textit{master initiator} Sub-C. Then, the \textit{master initiator} runs the \textit{new master} component \textit{B}. \textit{Master} component \textit{B} receives the list of registered \textit{actors} from \textit{master} component \textit{A} to advertise itself. After the initiation of \textit{master} component \textit{B}, it can also serve the placement requests of IoT \textit{Uers}.    

\subsection{Task Executor Component}
IoT applications can be separated into multiple dependent/independent \textit{task executor} containers based on the properties of the IoT application. Thus, an application can be easily deployed on several hosts for distributed execution. Moreover, \textit{task executors} can be efficiently reused for other requests of the same type, which significantly reduces the tasks' deployment time. To obtain this, when a \textit{task executor} finishes the execution of a specific \textit{user's} task, it goes into a cooling-off period. In this period, the container can be reused to serve another request.

\subsubsection{Executor}
The \textit{executor} Sub-C performs the run command to start the task. Also, it sends the results to the dependent children \textit{task executors} (in IoT applications with dependent tasks) or \textit{master} component (when there is no dependency).

\subsection{Remote Logger Component}
To support different application scenarios, this component can run on any hosts in edge/fog or cloud layers. All components send their periodical or event-driven logs to the \textit{Remote Logger}. This component collects the data and stores them in persistent storage, either using a file system or database. The \textit{Remote Logger} can connect to different databases distributed on any hosts, which enable IoT application scenarios that require distributed databases. In our current implementation, however, we run three databases in one host, including images (keeps the information about available docker images on different hosts), resources (keeps the information about hardware specifications of hosts), and system performance (keeps the information about response time, processing time, packet sizes, etc. of IoT applications). Moreover, the databases are containerized for faster deployments. Fig~\ref{fig:database_design} depicts an overview of databases and their tables. 

\subsubsection{Logger Manager}
The \textit{logger manager} Sub-C receives logs from \textit{masters}, \textit{actors}, and \textit{task executors} and keeps them in the persistent storage. For efficient and quick tracking of logs, the \textit{local manager} keeps the records of system performance, available resources, and containers' information on different storage. Also, \textit{logger manager} Sub-C can provide the latest logs of the system for the \textit{master} components. Besides, the stored logs can be used to analyze the overall status of the system.

\subsubsection{Database Design}
\begin{center}
    \begin{figure}[t]
    \centering 
    \includegraphics[width=\linewidth]{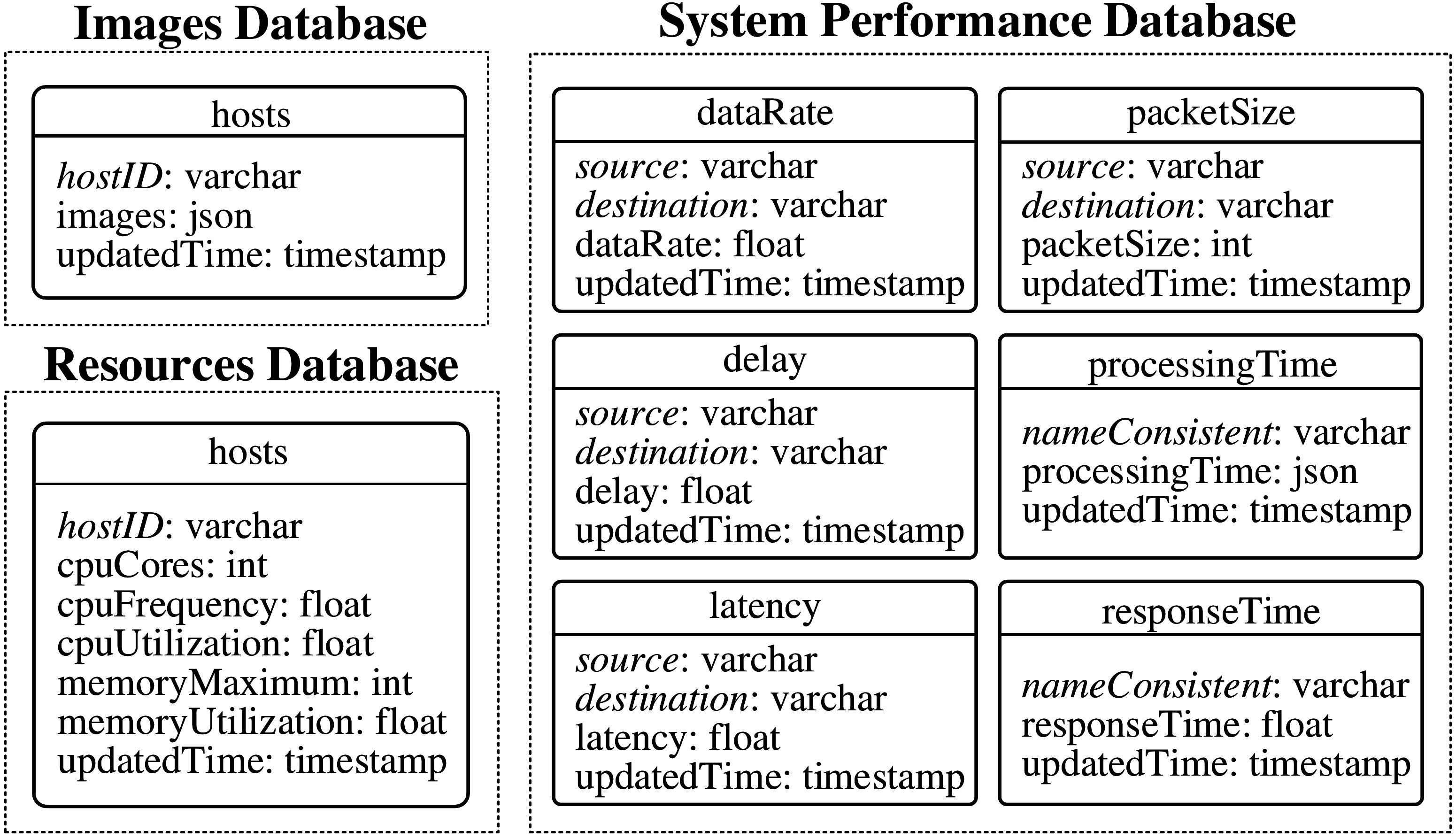}
    \caption{Database design}
    \label{fig:database_design}
\end{figure}

\end{center}

As Fig~\ref{fig:database_design} indicates, there are three main databases designed for the working of the FogBus2 framework. The Image Database contains the information of docker images on every host which have been seen by the framework. Image information is profiled by $actor$ periodically, and it is sent to $master$ within the registration message at the very beginning of the $actor$ registers. The image information will also be uploaded periodically to \textit{remote logger} to keep the information updated. Since $master$ synchronizes its logs with \textit{remote logger}, $master$ will ultimately get the image information of $actors$ over the network. Resources Database keeps the system's dynamic logs such as CPU cores, CPU frequency, CPU utilization, memory capacity, and memory utilization of every seen host. This information help scheduler and scaler work more reasonably, enabling the scheduler and scaler algorithms to find a better host for the workload. Compared with Resources Database and Image Database, System Performance Database is updated more frequently because it contains more dynamic information of the running system, including data rate, packet size, delay and latency between each host pairs, as well as processing time of \textit{task executor} for particular task and response time from $user$'s perspective. 

\section{Interaction Diagram}
\label{sec:userSequnce}
\begin{figure}[t]
    \centering 
    \includegraphics[width=\linewidth]{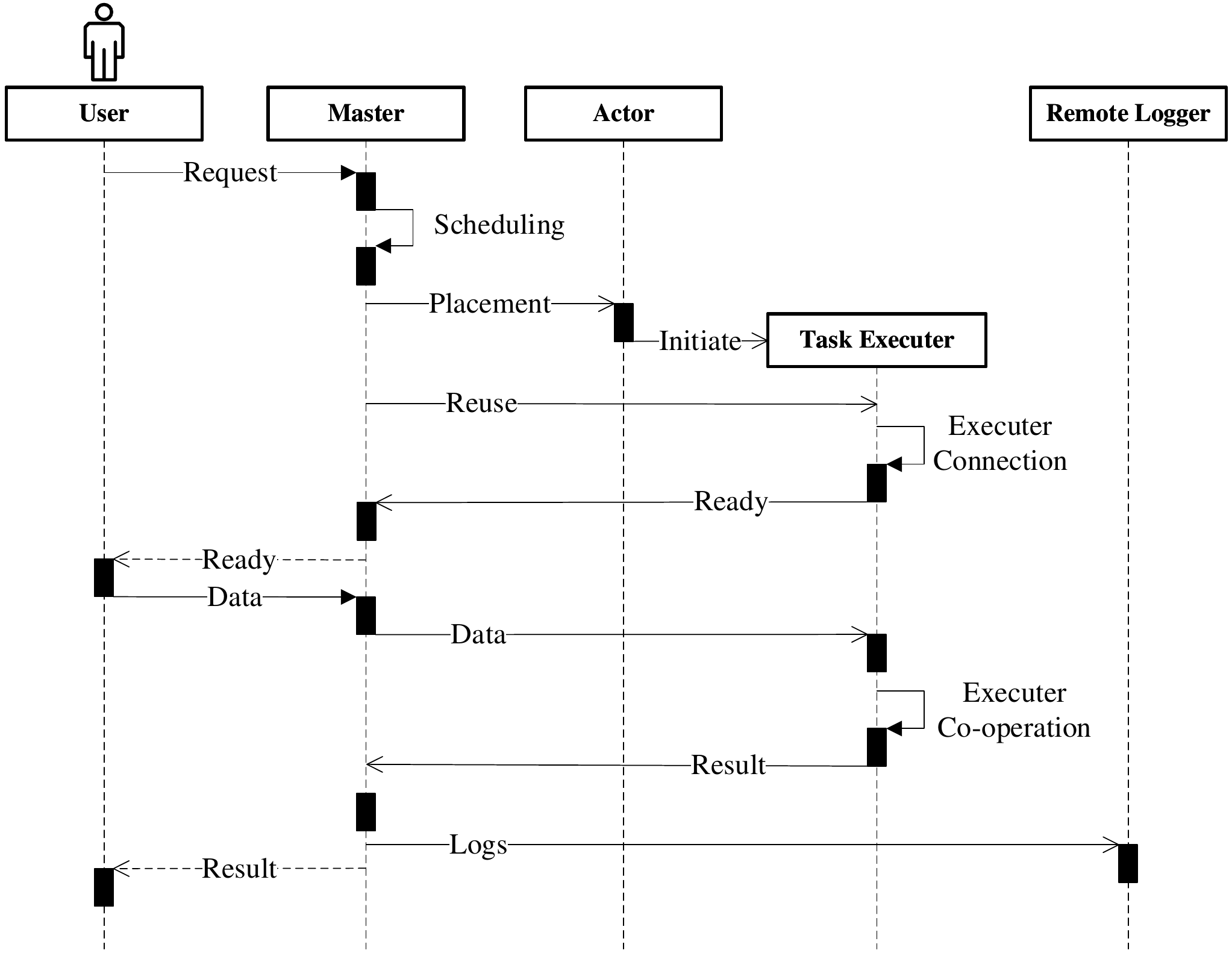}
    \caption{Interaction Diagram}
    \label{fig:userSequenceDiagram}
\end{figure}

We have discussed the five components in FogBus2, $user$, $master$, $actor$, \textit{task executor}, and \textit{remote logger}. Figure~\ref{fig:userSequenceDiagram} presents the  interaction diagram, which explains how an application request from $user$ is scheduled and what functionality will be invoked during the procedure.
The sequence begins with $user$'s request to $master$. After the $master$ receives the request from the $user$, it applies scheduling policies trying to decide how to arrange the execution. Once the scheduling algorithm finishes, the decision will be parsed to the respective $actors$ or \textit{task executor}. For $actors$, the placement messages are sent to initiated new \textit{task executor} containers. For \textit{task executor}, the reuse messages stop \textit{task executor}s' cool-off period and trigger \textit{task executor} to execute for the new placement. No matter the placement is initiating a new \textit{task executor} or reusing cool-off \textit{task executor}, the $task\; executor$ who received the placement message always connects the dependent peers based on which application is requested. The relationship is contained in the placement message. Next, a \textit{task executor} has to connect to its dependent peers, after which it acknowledges the $master$ with a message to indicate its ready state. If all the required \textit{task executor} are ready, $master$ then acknowledges $user$ that the resources needed for the requested application are all ready. Once the $user$ receives the ready message, it starts to send the data that need to be processed to the system and receives the response. The logs during the procedure will be uploaded to \textit{remote logger} for dynamical system performance monitoring, administrator maintenance, and further analysis.
To check the list of important messages, which are used in FogBus2 framework, refer to \cite{goudarzi2021resource}.

\chapter{Performance Evaluation}
\label{chapt:PerformanceEVAL}

This chapter discusses the properties of two sample container-based applications to represent real-time and non-real-time IoT applications. Also, we describe our experiments and evaluate the performance of the FogBus2 framework in real-world environments.

\section{Sample Container-based Applications}

\subsection{Conway's Game of Life}
It is a well-known 2D simulation game that consists of a grid of cells, where each cell can be either black or white. To obtain the next state of the grid, a local function must be applied to each cell simultaneously \cite{rendell2002turing}. In our implementation, each cell is defined as a pixel, and a group of pixels is defined as a rectangle. Our 2d world is separated into several rectangles of different sizes, incurring different computation sizes. Besides, these rectangles have a pyramid structure that defines a dependency model between different rectangles. Hence, we consider Conway's Game of Life as a real-time application with 62 dependent \textit{task executor} containers (one for each rectangle) with different computation sizes. 

\subsection{Video Optical Character Recognition (VOCR)} 
Compared to the pure OCR application, our implemented VOCR does not require any manual image input from users. The VOCR can either receive a live stream or pre-recorded video and automatically identify key-frames containing text. To filter key-frames, we used two different techniques, called Perceptual Image Hashing (pHash) and Hamming Distance. Then, for each keyframe, the text is extracted using the OCR technique. Finally, we apply the Editing Distance technique to filter the extracted texts which are similar. Our VOCR application can be used to extract text from books and important information about objects, such as objects in museums. We consider the VOCR as the non-real-time application in its current use-case since the text outputs are not required in real-time for users. However, the VOCR can also be used by smart vehicles in real-time scenarios such as reading traffic signs and warning messages on the road.

\section{Discussion on Experiments}
To study the performance of FogBus2 and its integrated policies, three experiments are conducted. In the first experiment, we analyze the scheduling mechanism of FogBus2 using different scheduling policies. Therefore, we integrate our proposed scheduling policy alongside two other policies in the FogBus2 framework. These policies attempt to approximate the real response time of IoT applications while considering different server configurations and find the best possible server configuration for the execution of IoT applications. Since all integrated scheduling policies are based on evolutionary algorithms, the estimated response time of IoT applications in different iterations is obtained to analyze the convergence rate of different scheduling policies. Moreover, we evaluate the real response time of IoT applications based on the obtained solutions from scheduling policies.

In the second experiment, we analyze the performance of the scalability mechanism of the FogBus2 framework. Typically, IoT integrates thousands and millions of devices that may send their requests to distributed \textit{master} components. These \textit{master} components are geographically distributed, and each one serves several IoT devices so that alongside other \textit{master} components, they can serve thousands or millions of IoT devices. So, in this experiment, IoT devices send a different number of simultaneous placement requests to each one of available \textit{master} components in the environments. Therefore, we study how efficiently the scalability mechanism of the FogBus2 framework can perform when the number of simultaneous requests to each \textit{master} component increases.

In the third experiment, we analyze the performance of the reuse mechanism of the FogBus2 framework. We normalized the resource reuse time by the VOCR without a reuse mechanism by comparing the time to evaluate the performance of the reuse mechanism for complex and straightforward dependent module applications.

In the fourth experiment, we analyze and compare the resource usage of our framework in terms of its startup time and RAM usage with its counterparts.

\section{Analysis of Scheduling Policies}
\label{sec:analysisOfSchedulingPolicies}
This experiment studies the performance of our proposed \textit{OHNSGA} scheduling algorithm and compares it with two other integrated scheduling policies in FogBus2, called Non-dominated Sorting Genetic Algorithm 2 (\textit{NSGA2}) as used in \cite{deb2002fast}, and Non-dominated Sorting Genetic Algorithm 3 (\textit{NSGA3}) \cite{deb2013evolutionary}. To keep fairness, the parameters of all scheduling policies are the same, including population size, maximum iteration number, and crossover probability. 

In this experiment, the environment contains 2 RPi 4B (ARM Cortex-A72 4 cores @1.5GHz CPU, and one with 2GB and another one with 4GB of RAM), and 1 Desktop (Intel Core i7 CPU @3.6GHz and 16 GB of RAM) to show the heterogeneity of servers in the edge layer. Also, the cloud layer contains 2 computing instances provisioned from Huawei Cloud (Intel Xeon 2 cores and 4 cores @2.6GHz CPU with 4GB and 8GB of RAM, respectively). The Desktop acts as a \textit{master} while it also can act as \textit{actor} to start \textit{tasks executors}. The rest of the hosts acts as \textit{actors} and runs \textit{task executors}. \textit{Master profiler} dynamically collects data about network characteristics of the environment (bandwidth and latency), IoT devices, and IoT applications. In this experiment, IoT devices send their requests for the execution of Conway's Game of Life application.

\begin{figure}[t]
    \centering 
    \includegraphics[width=\linewidth]{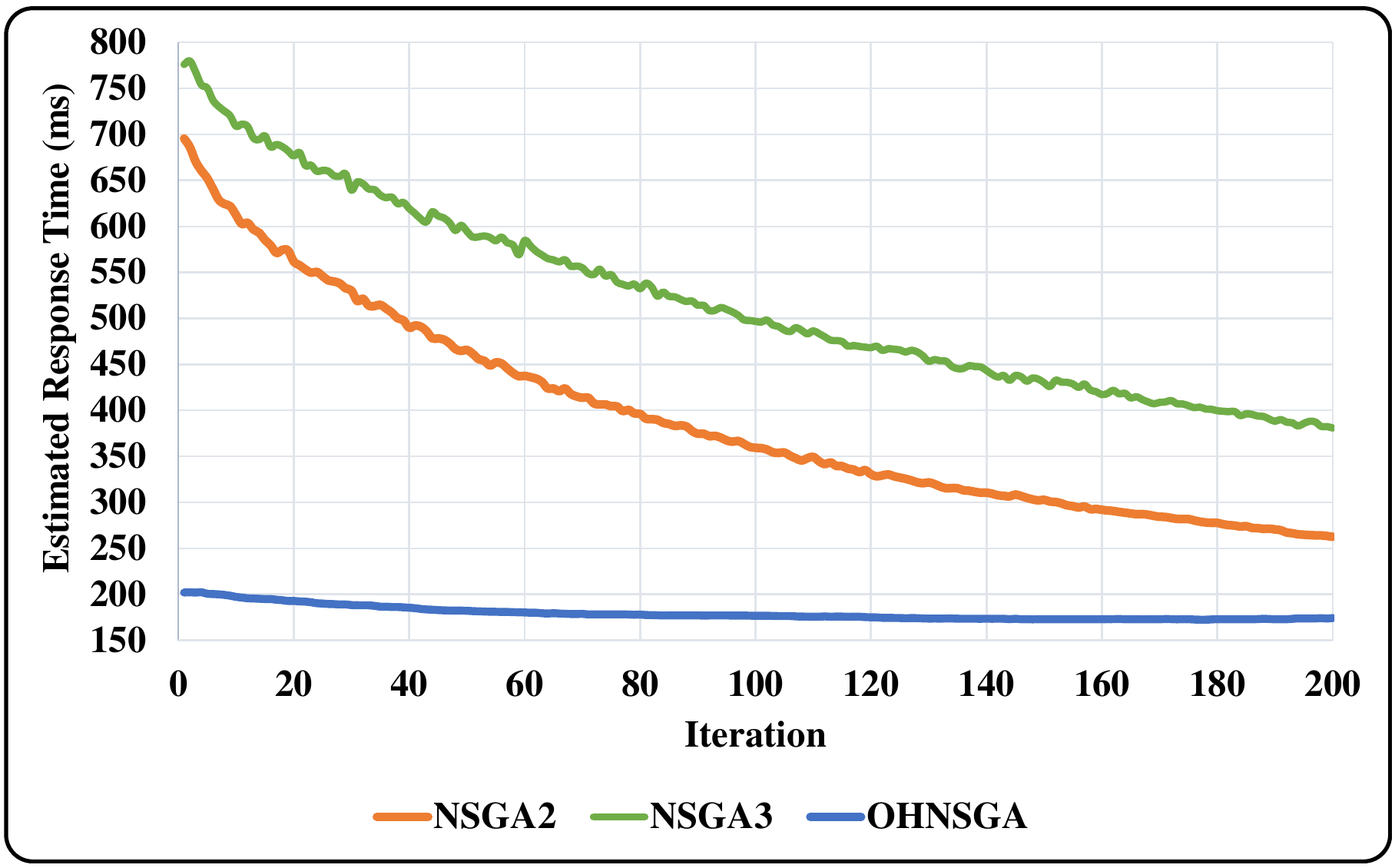}
    \caption{Scheduling performance in different iterations}
    \label{fig:experiment1_GA_iterations}
\end{figure} 
Fig.~\ref{fig:experiment1_GA_iterations} shows the average estimated response time of Conway's Game of Life application, obtained from different policies as the number of iterations increases. The \textit{OHNSGA} outperforms other policies and converges faster to better solutions. \textit{OHNSGA} keeps the records of previous decisions and profiling information for each application and initializes a part of its population using its recorded history. Besides, the optimized selection step of \textit{OHNSGA} ensures that non-duplicated best individuals can be copied to the next population. Therefore, \textit{OHNSGA} starts with better individuals compared to \textit{NSGA2} and \textit{NSGA3} due to its more intelligent initialization and keeps its diversity by selecting non-duplicated individuals for the next population. Accordingly, \textit{OHNSGA} can obtain faster convergence to better solutions in comparison to its counterparts.

\begin{figure}[t]
    \centering 
    \includegraphics[width=\linewidth] {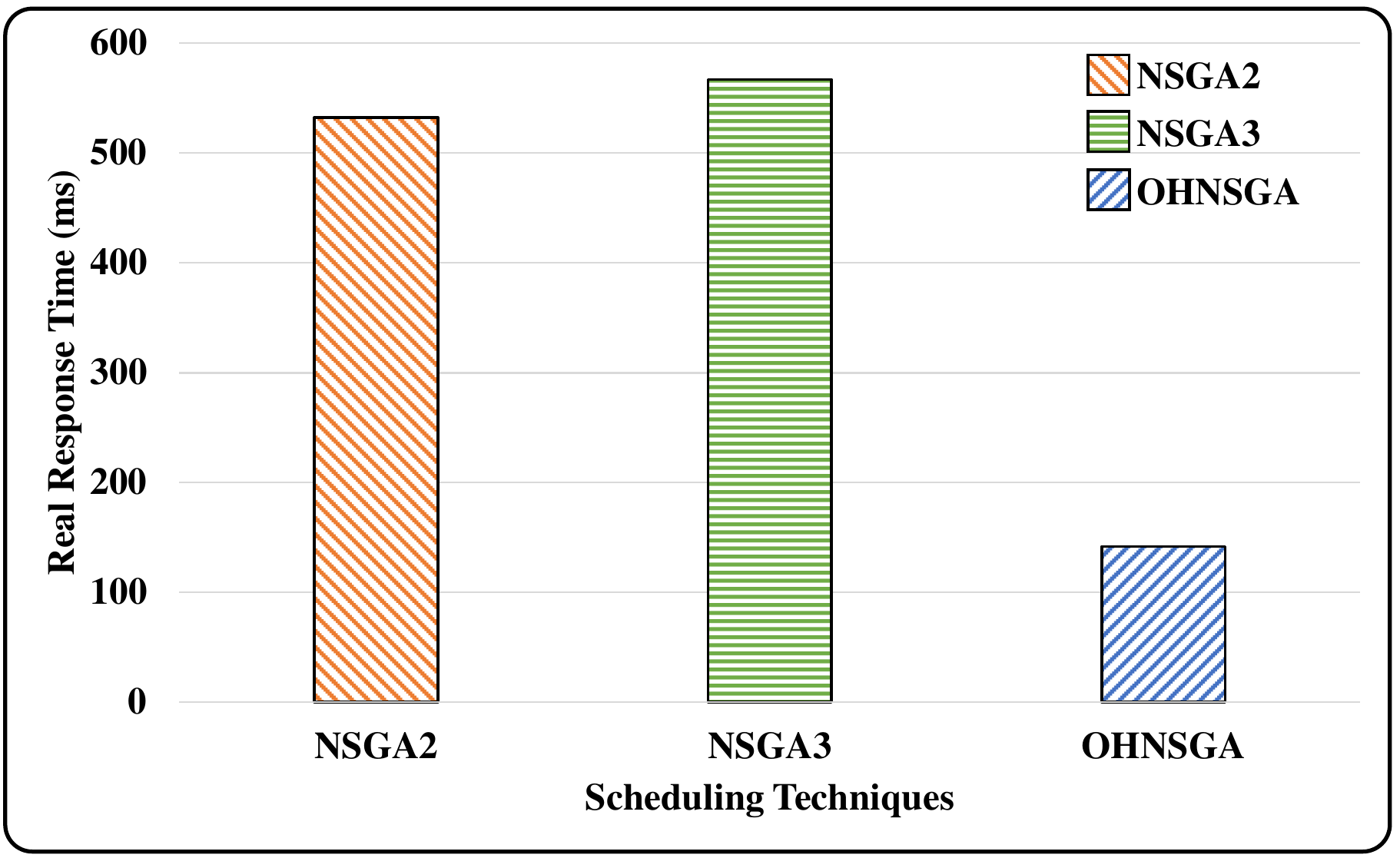}
    \caption{Real response time of scheduling policies}
    \label{fig:experiment1_real_estimated_compare}
\end{figure} 
Fig.~\ref{fig:experiment1_real_estimated_compare} depicts the real-world response time of Conway's Game of Life application, obtained from the execution of tasks in the real-world environment while considering different scheduling policies. As \textit{OHNSGA} tracks the prior execution behaviors of each application, its obtained real response time is less than other techniques. It proves that not only the \textit{OHNSGA} converge faster to a better solution compared to other policies, but its estimated solutions can better represent the behavior of the Game of Life in real-world environments.

\section{Analysis of Master Components' Scalability}
\label{sec:analysisOfMasterScalability}
In this experiment, the environment contains 4 RPi 4B (all with ARM Cortex-A72 4 cores @1.5GHz CPU, where two have 2GB RAM and the other two have 4GB RAM), 1 Desktop (Intel Core(TM) i7 CPU @3.6GHz and 16 GB of RAM) to represent the heterogeneity of servers in the edge layer. Moreover, the cloud layer contains five computing instances provisioned from Huawei Cloud (three instances with Intel Xeon 2 cores @2.6GHz CPU with 4 GB of RAM, and two instances with Intel Xeon 4 cores @2.6GHz CPU with 8 GB RAM). The \textit{master} and \textit{actors} are set as the same as in the previous experiment. Also, IoT devices send simultaneous requests of Conway's Game of Life and VOCR to the \textit{master}. We analyze two scenarios, called scalability and no-scalability. In the scalability scenario, the FogBus2's \textit{master} container scales up either when the number of received IoT requests increases or when the CPU utilization of the host on which the \textit{master} container is running goes above a threshold. The new \textit{master} container can be initiated on any host with sufficient resources, and the rest of the incoming requests can be managed by all available \textit{master} containers. In the no-scalability scenario, incoming requests to the \textit{master} container will be queued until enough resources for scheduling becomes available. Here, we define a Scheduling Finish Time (SFT) metric as the time difference when each IoT device sends its request to the \textit{master} until the \textit{master} container finishes the scheduling of the request. Hence, the SFT contains the queuing time of the request in the \textit{master} plus the scheduling time.

\begin{figure}[t]
    \centering 
    \includegraphics[width=\linewidth]{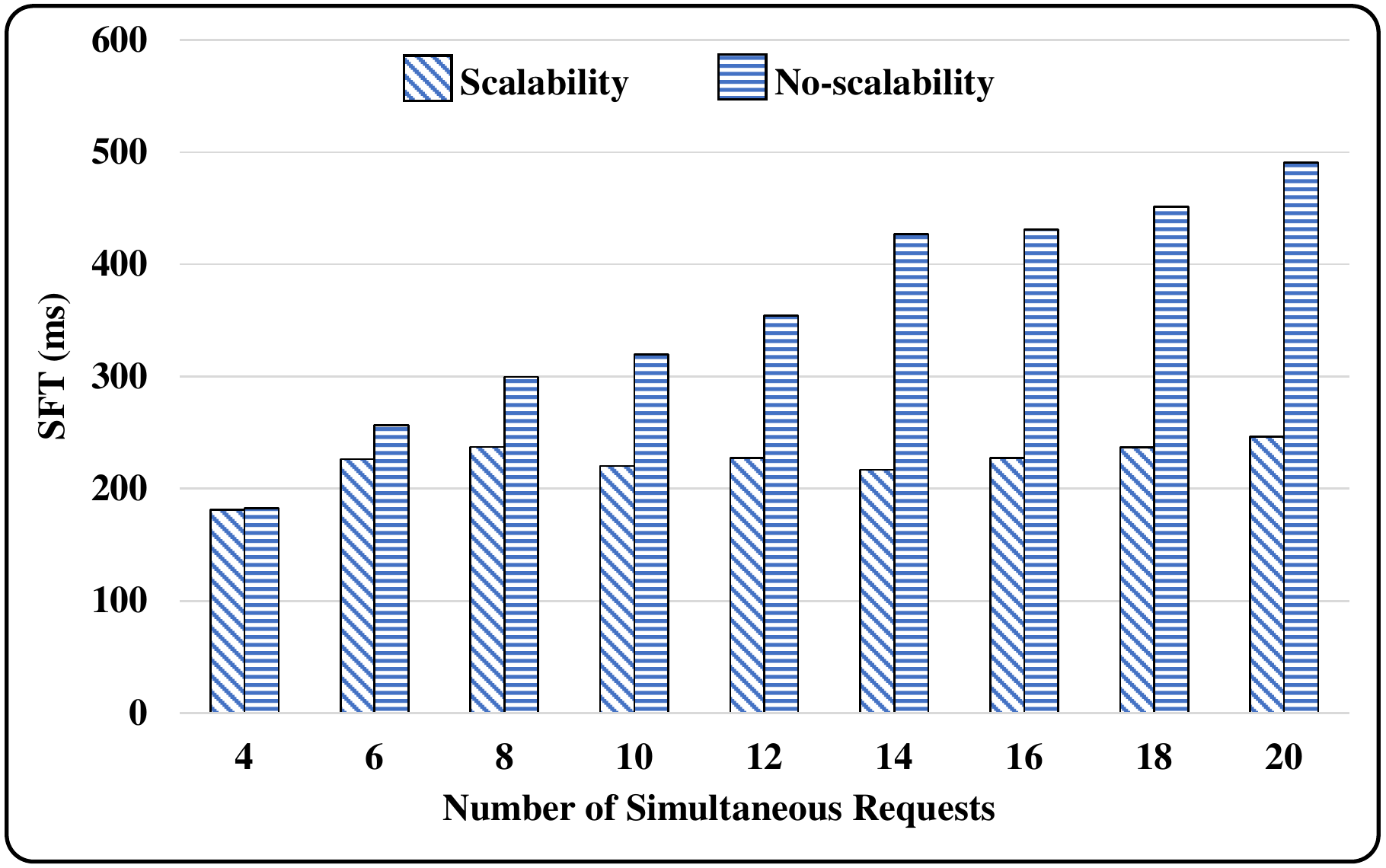}
    \caption{Analysis of master components' scalability}
    \label{fig:experiment2_scalability}
\end{figure}
Fig.~\ref{fig:experiment2_scalability} shows the scalability results as the number of simultaneous requests from IoT devices increases. The SFT values of both scenarios are roughly the same when the number of concurrent requests is small. However, as the number of requests increases, the SFT values of the no-scalability scheme dramatically increase compared to the scalability scenario. It shows the importance of supporting scalability mechanisms and policies in FogBus2. The \textit{master} containers are scaled up as the number of requests increases, which significantly reduces the queuing time of requests.    

\section{Analysis of Reusing Task Executor Components' Container}
\label{sec:taskExeutorReuse}
\begin{figure}[t]
    \centering 
    \includegraphics[width=\linewidth]{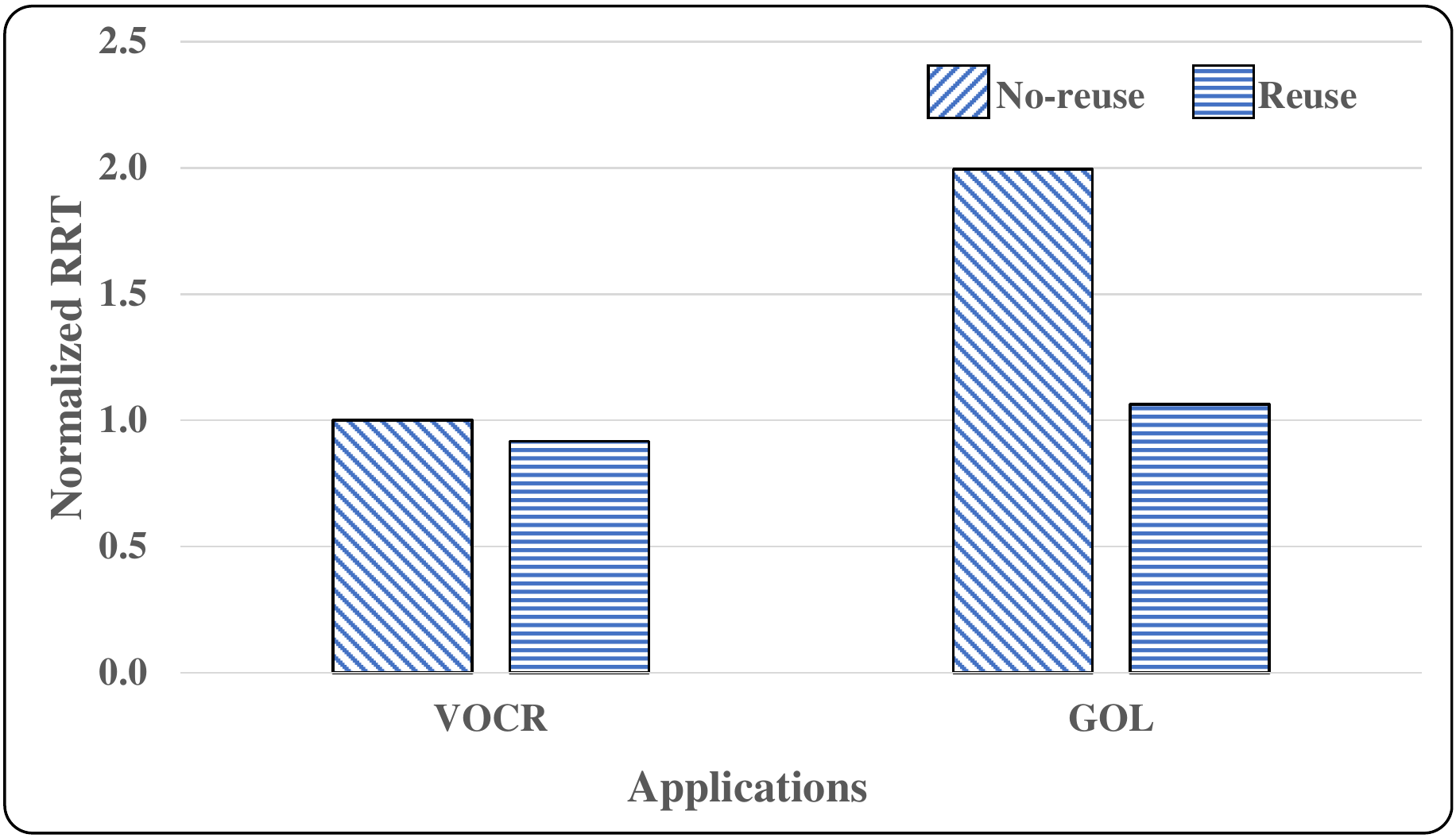}
    \caption{Analysis of reuse of task executor component; GOL is for Conway's Game of Life}
    \label{fig:experimentReuse}
\end{figure}
Fig.~\ref{fig:experimentReuse} presents the normalized resource ready time (RRT) of Conway's Game of Life and VOCR. RRT is considered from when $User$ sends the request to when it is informed that required resources are ready, i.e., \textit{task executors} are all ready. The experiment has been conducted on the Desktop only because we only care how much time the mechanism can save when with the reuse mechanism. The RRTs showing in figure~\ref{fig:experimentReuse} are normalized by the time of VOCR, which is without a reuse mechanism. For VOCR, a computation-intensive application, the reuse mechanism saves time but slightly. Because VOCR only requires few modules (tasks), i.e., containers, to be prepared. Compared with VOCR, Conway's Game of Life requires various more containers, dramatically raising the RRT without the reuse mechanism. However, when the reuse mechanism is used, even requiring resources to be placed and ready, the RRT of Conway's Game of Life decreases nearly 50 percent, almost equal to VOCR when VOCR requires just a few containers. It proves that the reuse mechanism runs efficiently, particularly for applications with complicated dependencies and relationships, because it avoids repeated container creation and initiation.

\section{Analysis of Startup Time and RAM Usage}
\label{sec:startupTimeAndRAMUsageAnalysis}
\begin{figure}[t]
    \centering 
    \includegraphics[width=\linewidth]{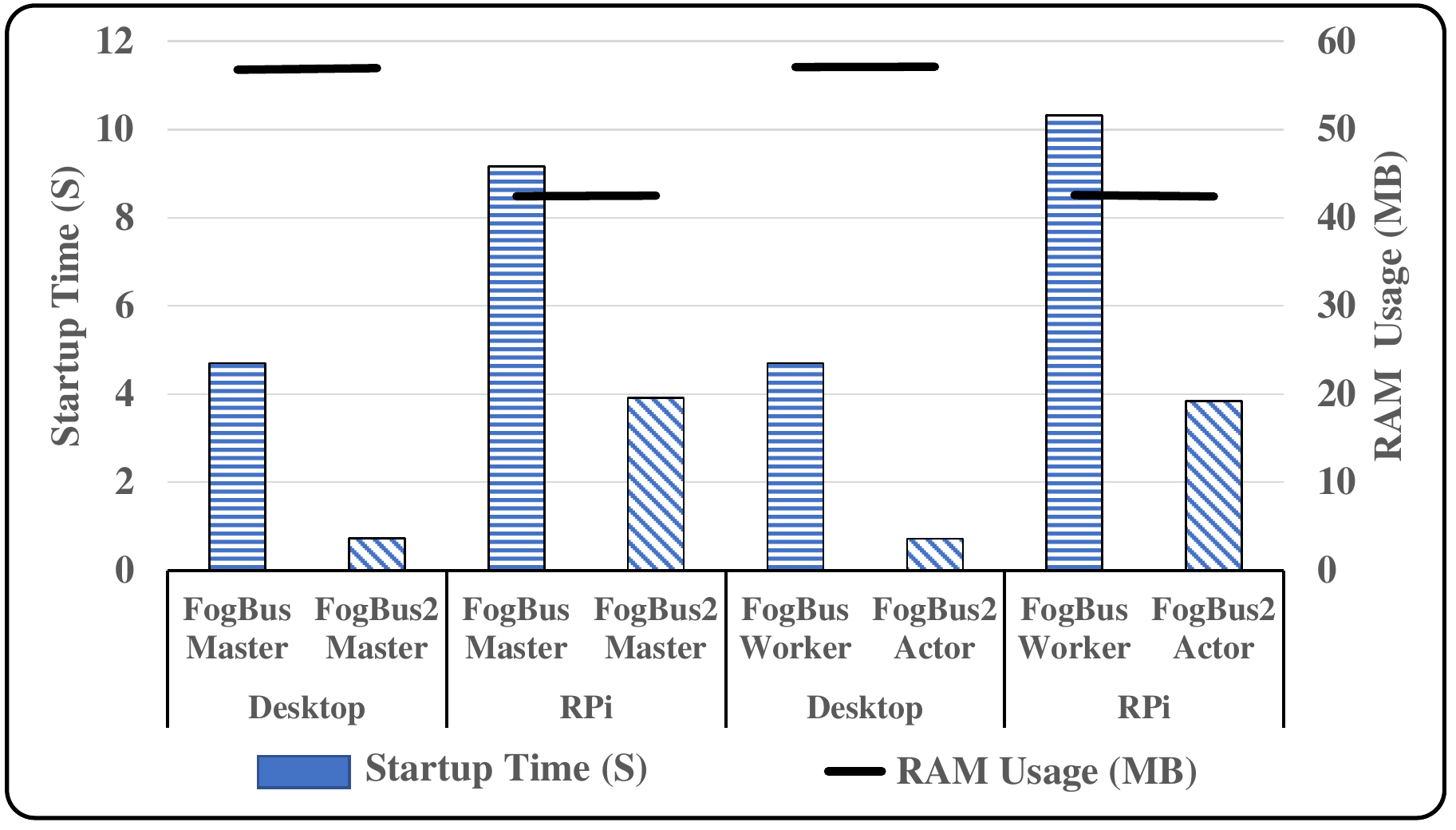}
    \caption{Startup time and RAM usage analysis}
    \label{fig:experiment3_startupRAMUsage}
\end{figure}
This experiment studies the startup time and RAM usage of our framework, FogBus2,  and compares it with FogBus framework~\cite{tuli2019fogbus}. Fig.~\ref{fig:experiment3_startupRAMUsage} shows the average startup time and RAM usage of \textit{master} and \textit{actor} components on different hosts. As the results are roughly the same for other components in our framework, we only present the obtained results for these two components. It can be seen the RAM usage of FogBus and our proposed framework, FogBus2, is roughly the same for different framework components. However, the startup time of FogBus2 is roughly 80\% and 60\% faster compared to FogBus on Desktop and RPi, which makes it a suitable option for fast deployment of any IoT-enabled systems.

\chapter{Conclusions and Future Directions}
\label{chapt:Conclusion}

\section{Conclusions}
In this work, we proposed FogBus2, a lightweight and distributed container-based framework to integrate heterogeneous IoT-enabled systems with edge/fog and cloud servers, with the following main contributions:
\begin{itemize}
    \item \textbf{FogBus2 offers fast and low-overhead deployments of applications using containerization.} By using the containerization technique, components of FogBus2 run in isolation environments on edge/fog servers and cloud servers. The runtime environments are secure and lightweight, which benefits FogBu2 and empowers FogBus2 with quick deployment. The required running environments of components are pre-compiled and packaged in images when the images can create as many containers as needed. Moreover, the containerization also decreases the difficulty of designing a scaling mechanism and scaling policies since the needed environments are already prepared in the images.
    \item \textbf{FogBus2 provides scheduling, scalability, resource discovery, and dynamic profiling mechanisms, assisting IoT developers in defining and deploying their targeted IoT applications on FogBus2.} We proposed $OHNSGA$ to dynamically schedule for heterogeneous requested IoT applications when $OHNSGA$ converges extremely fast compared with $NSGA2$ and $NSGA3$, and the decisions of $OHNSGA$ are examined to be better (53\% lower response time) in a real-world experiment. Besides, considering the workload growing into the system is unpredictable, an efficient mechanism to automatically scale resources has to be developed when the investigation shows a 48\% improvement compared with no scalability is applied. Moreover, the resource discovery mechanism and dynamic profiling mechanism are integrated into FogBus2 to locate undefined resources and monitor system performance automatically. 
    \item \textbf{FogBus2 does not have any constraints on communication topology between its entities and supports different topologies such as mesh, peer-to-peer, and client-server.} The communication of every components in FogBus2 framework design, $master$, $actor$, $user$, \textit{task executor}, and \textit{remote logger}, do not couple with others. For example, \textit{task executor}s can communicate to each other without the participation of $master$, which dramatically increases the efficiency for \textit{task executor}s to co-operate with others when the computation requires content exchanging. This loose design of communication enables various communication topology and make FogBus2 more compatible with different networking environment. It also gives developers more freedom the develop their applications over our framework and contribute to FogBus2.
\end{itemize}

\section{Future Directions}
Due to modular design and containerization support, IoT developers can easily extend this framework and integrate new software components and policies. Hence, this framework can be further developed by,
\begin{itemize}
    \item \textbf{Integrating dynamic clustering mechanisms and policies to cluster resources either horizontally or vertically.} When seamless integration of edge/fog and cloud infrastructures has been developed in FogBus2 to support heterogeneous IoT applications, a horizontal and vertical dynamic clustering mechanism can also be integrated upon the integration capability. To intelligently cluster highly heterogeneous resources, a real-world mechanism may be developed learning from simulations like the mechanism proposed in \cite{kumar2009eehc}.
    \item \textbf{Integrating container-orchestration techniques to automate the management of application deployments and scaling.} FogBus2 manages the containers of different components and applications using the Docker API with the developed algorithm of the scaler. However, automatic management techniques to deploy and scale containers, like Kubernetes \cite{bernstein2014containers}, can be integrated to obtain more efficient practical performance.
    \item \textbf{Mobility-support in different layers of edge/fog computing environment.} Since IoT devices are usually tiny and some of them are mobile such as in-vehicle cameras, the mobility support for IoT users and edge/fog servers to provide users seamless and stable experience is also essential. The potential research may refer to Sufyan et al. \cite{almajali2018framework}.
    \item \textbf{Integrating lightweight security mechanisms to ensure data confidentiality and integrity} Integration of security mechanisms usually requires extra overhead, but there are several works \cite{HASSAN2019512,tuli2019fogbus,7917634,8306880,7946872} that can be referred to integrate blockchain technique which minimizes the overhead on security monitoring and also keeps tracking the sensitive information during the execution of applications and serving of the framework.
    \item \textbf{Privacy preservation support for the users' private information and edge/fog servers.} Privacy preservation is significant for IoT applications when the IoT devices are close to the natural environment and to humans. For example, healthcare applications highly concern about privacy and security because the applications service and run with patients' sensitive data. When the transmission of such data is over the edge networking, the protection and preservation need to be considered, which can be referred to Bakkiam et al. \cite{8839043} and Sahi et al. \cite{8089328}.
    \item \textbf{Integrating machine learning techniques to analyze the current state of edge/fog computing environment.} It is difficult to understand and efficiently manage an edge/fog computing environment because the current state of such distributed system is incredibly dynamic and complex. Referring to \cite{TULI2020187,tuli2020dynamic,GOUDARZI2019102407,8603156}, the machine learning techniques can be integrated into the FogBus2 framework to understand and analyze the system, as a consequence, develop and improve policies for scheduling, scaling, and resource discovery mechanism.
\end{itemize}







\addtocontents{toc}{\vspace{2em}}  
\backmatter

\label{Bibliography}
\lhead{\emph{Bibliography}}  
\bibliographystyle{unsrtnat}  
\bibliography{Bibliography}  

\end{document}